\documentclass[aps,prb,times,twocolumn,amsmath,amssymb,superscriptaddress,floatfix]{revtex4}

\usepackage{color}

\usepackage{graphicx}
\usepackage{subfigure}
\usepackage{bbold}
\usepackage{verbatim}
\usepackage{float}
\usepackage{enumerate}
\usepackage{amsfonts}
\usepackage{multirow}
\usepackage[colorlinks,bookmarks=false,citecolor=blue,linkcolor=red,urlcolor=blue]{hyperref}

\newcommand{\dg}{^\dagger}
\newcommand{\pdg}{^{\phantom\dagger}}

\begin{document}


\title{
Magnon binding in BaCdVO(PO$_4$)$_2$
}


\author{Andrew Smerald}
\email{andrew.smerald@gmail.com}

\affiliation{Max Planck Institut f\"ur Festk\"orperforschung, Heisenbergstra\ss e 1, D-70569 Stuttgart, Germany}


\date{\today}


\begin{abstract}  
The bond-nematic state has both long-range ordering of magnetic quadrupoles and the high entanglement typical of spin liquids, and as such can be thought of as a spin liquid crystal.
One of the most promising materials in which to find such a state is BaCdVO(PO$_4$)$_2$, in which a magnetically silent phase has been found between a magnetically ordered low-field phase and the high-field polarised phase.
Here I study the magnetic Hamiltonian of BaCdVO(PO$_4$)$_2$, and, for the Hamiltonian-parameters determined by fits to inelastic neutron scattering experiments, show that magnons condense out of the polarised state as bound pairs.
This is direct theoretical support for the existence of a bond-nematic state just below the saturation field.
\end{abstract}

\maketitle


\section{Introduction}


There are many different flavours of spin-nematic order in condensed matter physics, and one of the most interesting is undoubtedly the bond nematic \cite{andreev84,chubukov87,chubukov91,shannon06,momoi06}.
In a bond nematic the spins perform an entangled dance that results in a multipolar order parameter forming on the bonds of the lattice, while at the same time the system has properties similar to a spin liquid \cite{shindou09,shindou11,momoi12,shindou13}.
In analogy with classical liquid crystals, the bond nematic state can be thought of as a spin liquid crystal.

Just like spin-liquid states, the stability of spin liquid crystals typically relies on strong frustration, and is fragile to additional magnetic interactions that relieve this frustration \cite{shannon06,momoi06,sindzingre09,sindzingre10}.
However, unlike spin-liquids, the bond nematic is built from triplet pairing of neighbouring spins, and it follows that its stability can be enhanced by magnetic field \cite{shindou09}.
In fact, the application of magnetic field can massively enhance the parameter window in which the bond-nematic state is stable, and this effect is most pronounced just below the transition into the fully polarised state \cite{shannon06,ueda09,ueda13,ueda15,smerald15}.

There are several candidate materials for the realisation of a bond-nematic phase \cite{svistov10,janson16,nawa17,orlova17,grafe17}, with one of the most promising being BaCdVO(PO$_4$)$_2$ \cite{nath08,povarov19,skoulatos19,bhartiya19}.
Intriguingly, as the field applied to BaCdVO(PO$_4$)$_2$ is lowered through the saturation value of approximately 6.5T, there is an apparently continuous phase transition from the fully-polarised state to a  partially-polarised but otherwise magnetically-silent state \cite{povarov19,skoulatos19,bhartiya19}.
This phase persists down to approximately 4T, where there is a transition to a low-field magnetically-ordered state.
At the same time the material is strongly frustrated, with competition between ferromagnetic nearest-neighbour exchange interactions and antiferromagnetic second-neighbour exchange \cite{nath08,povarov19,skoulatos19,bhartiya19}.
Taken together, this is suggestive of the formation of a bond-nematic state in a field window of approximately 4-6.5T, but not conclusive proof.

Strong evidence for the existence of a bond nematic would involve showing that a finite density of bound magnon pairs develops below the saturation field.
Condensation of such pairs out of the fully-polarised phase results in $\langle S_i^-S_j^- \rangle$ taking a finite value, which is exactly the bond-nematic order parameter \cite{shannon06,ueda13,smerald15}.

In this paper I study magnon binding in the magnetic Hamiltonian believed on symmetry grounds to describe BaCdVO(PO$_4$)$_2$.
I show that above the saturation field there is wide range of parameters in which a band of magnon bound states lie below the two-magnon continuum, and that this includes the parameters extracted from fits to high-field inelastic neutron scattering measurements \cite{bhartiya19}.
Crucially, I further show that condensation of these magnon pairs is the first instability of the fully-polarised phase on lowering magnetic field. 
Thus I provide direct theoretical support to the idea that BaCdVO(PO$_4$)$_2$ has a bond-nematic state below the saturation field.


\section{Magnetic Hamiltonian}
\label{Sec:Hamiltonian}

The minimal magnetic Hamiltonian of BaCdVO(PO$_4$)$_2$ consists of Heisenberg couplings between first and second neighbour spin-1/2's.
The spins are associated with V$^{4+}$ ions that approximately form square planes.
Interactions between spins in neighbouring planes are believed to be negligibly small compared to the in-plane interactions, and anisotropic interactions can be discounted due to the isotropic low-temperature g-factor and small spin-flop field \cite{povarov19,skoulatos19,bhartiya19}.

\begin{figure}[t]
\centering
\includegraphics[width=0.35\textwidth]{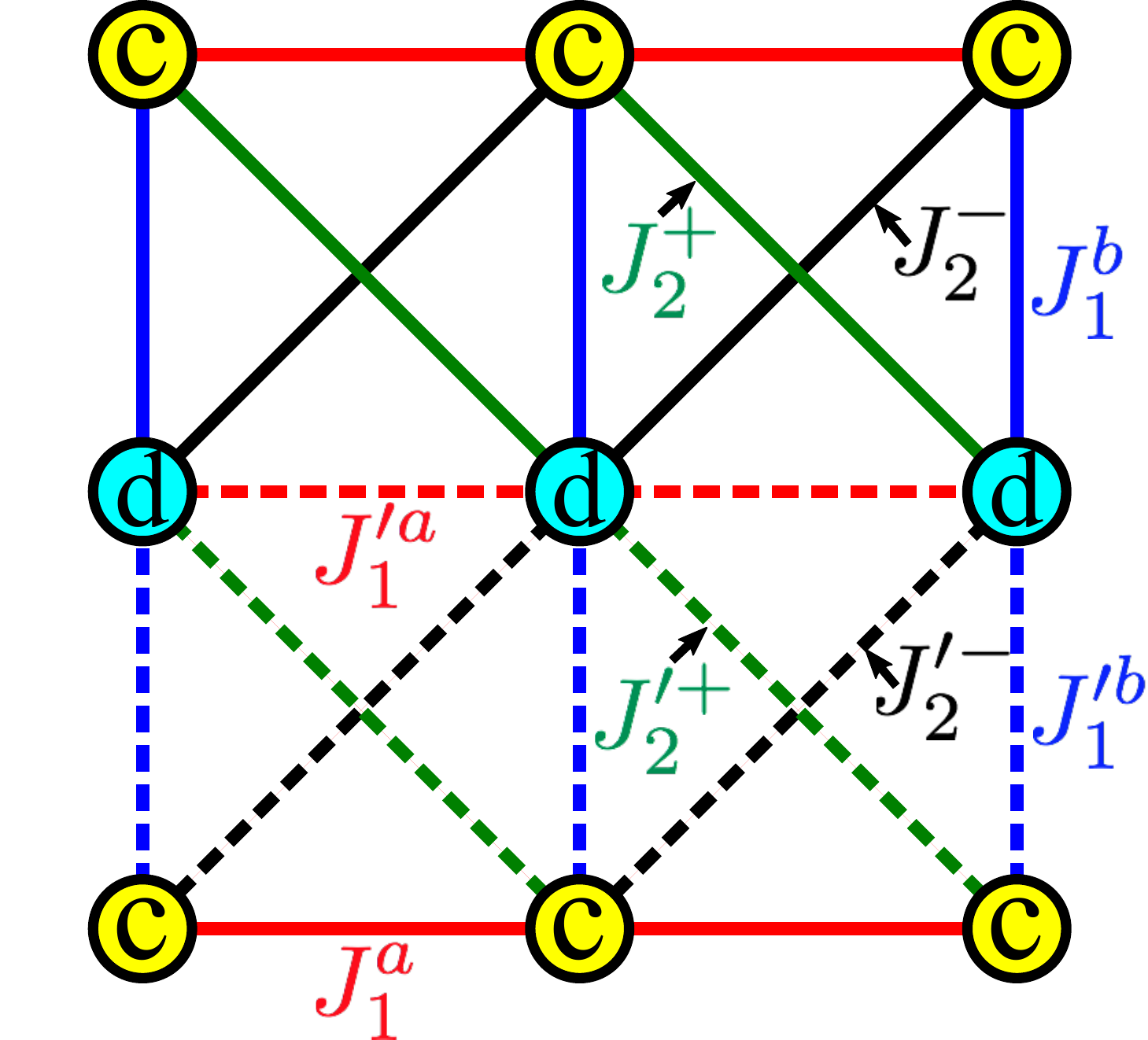}
\caption{\footnotesize{
Multi-$J$ Hamiltonian believed to describe the material BaCdVO(PO$_4$)$_2$.
There are 8 different magnetic couplings linking first and second neighbours.
As a result there are 2 inequivalent sites, which are labelled $c$ and $d$.
}}
\label{fig:J_interactions}
\end{figure}

Taking into account the small deviation of the V ions from a perfect square lattice, the most general Heisenberg Hamiltonian is given by \cite{bhartiya19},
\begin{align}
\mathcal{H}_{\sf spin} = 
& \sum_{\langle ij \rangle} J_{ij} {\bf S}_i \cdot {\bf S}_j
-h\sum_i S_i^{\sf z},
\label{eq:Hspin}
\end{align}
where ${\langle ij \rangle}$ runs over all first and second neighbour bonds, ${\bf S}$ is a spin-1/2 operator and there are eight distinct values of $J_{ij}$, shown in Fig.~\ref{fig:J_interactions}.
These $J_{ij}$ values have been parametrised from neutron scattering experiments, where good fits to the data were obtained for $J_1^a = J_1^{\prime a} = J_1^b = -0.42$ meV, $J_1^{\prime b} = -0.34$ meV, $J_2^+ = J_2^- = 0.16$ meV and $J_2^{\prime +} = J_2^{\prime -} = 0.38$ meV \cite{bhartiya19}.
While these parameters may be improved upon as further experiments are performed, they are consistent with the low-field magnetic-ordered state \cite{skoulatos19} and with the finding that there is competition between ferro and antiferromagnetism \cite{nath08}.
Since the magnetic interactions are relatively weak, the magnetic field required to saturate the system is also relatively weak, and the fully polarised state is realised above a field of about 6.5T \cite{bhartiya19}.

A useful starting point is to first rewrite $\mathcal{H}_{\sf spin}$ [Eq.~\ref{eq:Hspin}] in a bosonic form \cite{batyev85}.
Here I will do this under the assumption that $J_1^a = J_1^{\prime a}$, since this is consistent with the suggested parameters for BaCdVO(PO$_4$)$_2$ and also considerably simplifies all the mathematical expressions.
For completeness the case of  $J_1^a \neq J_1^{\prime a}$ is presented in Appendix~\ref{App:bosonH}.
Also, when considering geometrical factors, small deviations from a square lattice are ignored, due to their irrelevantly small size.

Since for the fully-polarised state there are two sites in the unit cell, which can be labelled $c$ and $d$ (see Fig.~\ref{fig:J_interactions}), it is necessary to introduce two distinct boson operators, according to,
 \begin{align}
i\in c \quad S_i^{\sf z} &= 1/2 - c_i\dg c_i, \quad
S_i^+ =  c_i, \quad
S_i^- =  c_i\dg , \nonumber \\
i\in d \quad S_i^{\sf z} &= 1/2 - d_i\dg d_i, \quad
S_i^+ =  d_i, \quad
S_i^- =  d_i\dg.
\end{align}
These transformations are exact, as long as a hardcore constraint is imposed on every site.
In consequence $\mathcal{H}_{\sf spin}$ can be rewritten in terms of a two-boson and a four-boson term as $\mathcal{H}= \mathcal{H}_2 + \mathcal{H}_4$.
The two-boson part is given by,
\begin{align}
\mathcal{H}_2= \sum_{{\bf q}}
\left(  c\dg_{\bf q},d_{\bf q}\dg \right)
\left(
\begin{array}{cc}
A_{\bf q} &B_{\bf q}  \\
B_{\bf q}^* &A^\prime_{\bf q}   
\end{array}
\right) 
\left(
\begin{array}{c}
c_{\bf q}\pdg \\
d_{\bf q}\pdg
\end{array}
\right) 
\label{eq:H2cd},
\end{align}
where ${\bf q}$ lives in the Brillouin zone of the 2-site unit cell ($-\pi<q_a<\pi$, $-\pi/2<q_b<\pi/2$), the lattice constant is set to unity and,
\begin{align}
A_{\bf q} &= h - J_1^a (1-\cos q_a) - \frac{1}{2}(J_1^b + J_1^{\prime b}) \nonumber \\
&- \frac{1}{2}(J_2^+ + J_2^- + J_2^{\prime +} + J_2^{\prime -}) \nonumber \\
B_{\bf q}  &= \frac{1}{2}(J_1^b e^{iq_b} + J_1^{\prime b} e^{-iq_b}) 
+\frac{1}{2}(J_2^+ e^{-i(q_a-q_b)}  \nonumber \\
& + J_2^- e^{i(q_a+q_b)} 
+ J_2^{\prime +} e^{i(q_a-q_b)} 
+ J_2^{\prime -} e^{-i(q_a+q_b)}).
\label{eq:AB}
\end{align}
Diagonalisation of $\mathcal{H}_2$ is achieved by introducing the bosons $\alpha$ and $\beta$ according to,
\begin{align}
\left(
\begin{array}{c}
c_{\bf q}\pdg \\
d_{\bf q}\pdg
\end{array}
\right) 
=
\left(
\begin{array}{cc}
u_{\bf q}^x  & v_{\bf q}^x  \\
u_{\bf q}^y  & v_{\bf q}^y    
\end{array}
\right) 
\left(
\begin{array}{c}
\alpha_{\bf q}\pdg \\
\beta_{\bf q}\pdg
\end{array}
\right) ,
\end{align}
where ${\bf u}_{\bf q}=(u_{\bf q}^x,u_{\bf q}^y)$ and ${\bf v}_{\bf q}=(v_{\bf q}^x,v_{\bf q}^y)$ are the eigenvectors of $\mathcal{H}_2$ and are given by,
 \begin{align}
{\bf u}_{\bf q} = \frac{1}{\sqrt{2}} \left(
\frac{B_{\bf q}}{|B_{\bf q}|},1
\right), \quad
{\bf v}_{\bf q} = \frac{1}{\sqrt{2}} \left(
-\frac{B_{\bf q}}{|B_{\bf q}|},1
\right).
\end{align}
The normalised and orthogonal nature of these eigenvectors is sufficient that $\alpha$ and $\beta$ inherit the proper boson commutation relationships.
The resulting Hamiltonian is,
\begin{align}
\mathcal{H}_2 = \sum_{\bf q} \left[ \omega_{\bf q}^\alpha \alpha\dg_{\bf q}\alpha_{\bf q} + \omega_{\bf q}^\beta \beta\dg_{\bf q}\beta_{\bf q}  \right],  \quad
 \omega_{\bf q}^{\alpha/\beta} = A_{\bf q} \pm |B_{\bf q}|.
 \label{eq:H2ab}
\end{align}

The 4-boson part of the Hamiltonian can be written as,
\begin{align}
\mathcal{H}_4 = \frac{1}{N} \sum_{{\bf k}_1...{\bf k}_4 } &\left[
\tilde{V}^{cc}_{{\bf k}_2-{\bf k}_4} c\dg_{{\bf k}_1} c\dg_{{\bf k}_2} c_{{\bf k}_3} c_{{\bf k}_4}
+ V^{cd}_{{\bf k}_2-{\bf k}_4} c\dg_{{\bf k}_1} d\dg_{{\bf k}_2} c_{{\bf k}_3} d_{{\bf k}_4} \right. \nonumber \\
&\left. +\tilde{V}^{dd}_{{\bf k}_2-{\bf k}_4} d\dg_{{\bf k}_1} d\dg_{{\bf k}_2} d_{{\bf k}_3} d_{{\bf k}_4}
\right] \delta_{{\bf k}_1+{\bf k}_2-{\bf k}_3-{\bf k}_4},
\label{eq:Vcd}
\end{align}
where $N$ is the total number of square lattice sites,
 \begin{align}
\tilde{V}^{cc}_{\bf q} = \tilde{V}^{dd}_{\bf q} = 2U+2J_1^a \cos q_a, \quad
V^{cd}_{\bf q} = 4B_{\bf q}, 
\end{align}
and $U$ is an infinite on-site potential that enforces the hardcore constraint.
Transforming to the boson operators that diagonalise $\mathcal{H}_2$  and rewriting in terms of the centre of momentum coordinates ${\bf k}_1= {\bf K}/2+{\bf p}$, ${\bf k}_2= {\bf K}/2-{\bf p}$, ${\bf k}_3= {\bf K}/2+{\bf p}^\prime$ and ${\bf k}_4= {\bf K}/2-{\bf p}^\prime$ gives,
\begin{align}
\mathcal{H}_4 = \frac{1}{N} \sum_{{\bf K}, {\bf p}, {\bf p}^\prime } &\left[
\tilde{V}^{\alpha\alpha\alpha\alpha}_{{\bf K},{\bf p},{\bf p}^\prime}  \alpha\dg_{{\bf K}/2+{\bf p}} \alpha\dg_{{\bf K}/2-{\bf p}} \alpha_{{\bf K}/2+{\bf p}^\prime} \alpha_{{\bf K}/2-{\bf p}^\prime} \right. \nonumber \\
&+\tilde{V}^{\alpha\beta\alpha\alpha}_{{\bf K},{\bf p},{\bf p}^\prime}  \alpha\dg_{{\bf K}/2+{\bf p}} \beta\dg_{{\bf K}/2-{\bf p}} \alpha_{{\bf K}/2+{\bf p}^\prime} \alpha_{{\bf K}/2-{\bf p}^\prime} \nonumber \\
&+\tilde{V}^{\alpha\alpha\alpha\beta}_{{\bf K},{\bf p},{\bf p}^\prime}  \alpha\dg_{{\bf K}/2+{\bf p}} \alpha\dg_{{\bf K}/2-{\bf p}} \alpha_{{\bf K}/2+{\bf p}^\prime} \beta_{{\bf K}/2-{\bf p}^\prime} \nonumber \\
&+\tilde{V}^{\alpha\alpha\beta\beta}_{{\bf K},{\bf p},{\bf p}^\prime}  \alpha\dg_{{\bf K}/2+{\bf p}} \alpha\dg_{{\bf K}/2-{\bf p}} \beta_{{\bf K}/2+{\bf p}^\prime} \beta_{{\bf K}/2-{\bf p}^\prime} \nonumber \\
&+\tilde{V}^{\beta\beta\alpha\alpha}_{{\bf K},{\bf p},{\bf p}^\prime}  \beta\dg_{{\bf K}/2+{\bf p}} \beta\dg_{{\bf K}/2-{\bf p}} \alpha_{{\bf K}/2+{\bf p}^\prime} \alpha_{{\bf K}/2-{\bf p}^\prime} \nonumber \\
&+\tilde{V}^{\alpha\beta\alpha\beta}_{{\bf K},{\bf p},{\bf p}^\prime}  \alpha\dg_{{\bf K}/2+{\bf p}} \beta\dg_{{\bf K}/2-{\bf p}} \alpha_{{\bf K}/2+{\bf p}^\prime} \beta_{{\bf K}/2-{\bf p}^\prime} \nonumber \\
&+\tilde{V}^{\alpha\beta\beta\beta}_{{\bf K},{\bf p},{\bf p}^\prime}  \alpha\dg_{{\bf K}/2+{\bf p}} \beta\dg_{{\bf K}/2-{\bf p}} \beta_{{\bf K}/2+{\bf p}^\prime} \beta_{{\bf K}/2-{\bf p}^\prime} \nonumber \\
&+\tilde{V}^{\beta\beta\alpha\beta}_{{\bf K},{\bf p},{\bf p}^\prime}  \beta\dg_{{\bf K}/2+{\bf p}} \beta\dg_{{\bf K}/2-{\bf p}} \alpha_{{\bf K}/2+{\bf p}^\prime} \beta_{{\bf K}/2-{\bf p}^\prime} \nonumber \\
 & \left.+ \tilde{V}^{\beta\beta\beta\beta}_{{\bf K},{\bf p},{\bf p}^\prime}  \beta\dg_{{\bf K}/2+{\bf p}} \beta\dg_{{\bf K}/2-{\bf p}} \beta_{{\bf K}/2+{\bf p}^\prime} \beta_{{\bf K}/2-{\bf p}^\prime} \right].
 \label{eq:H4ab}
\end{align}
The expressions for the $V$'s are quite lengthy, and are given in Eq.~\ref{eq:V}.


\section{Single magnon physics}


Before considering the binding of magnons, it is useful to first consider the single-magnon excitations of the fully polarised state.
These can be understood exactly, since hopping of bosons is purely due to $\mathcal{H}_2$, and is not affected by $\mathcal{H}_4$.

The single magnon dispersion has two branches, as would be expected for a 2-site unit cell, and the dispersion relationships are given by $ \omega_{\bf q}^{\alpha/\beta}$ [Eq.~\ref{eq:H2ab}].
The associated parameter space is large, and, rather than trying to study the excitations for all possible parameters, it is useful to instead consider a representative sub-space.
In order to do this one can define,
 \begin{align}
\tilde{J} = J_1^{ b} + J_2^{ +} + J_2^{ -}, \quad
\tilde{J}^\prime = J_1^{\prime b} + J_2^{\prime +} + J_2^{\prime -}.
\end{align}
The usefulness of these combinations comes from the fact that they are important for determining the minimum of $\omega_{\bf q}^{\alpha/\beta}$, and therefore which single-magnon mode is the first to go soft as the magnetic field is reduced.

\begin{figure}[t]
\centering
\includegraphics[width=0.4\textwidth]{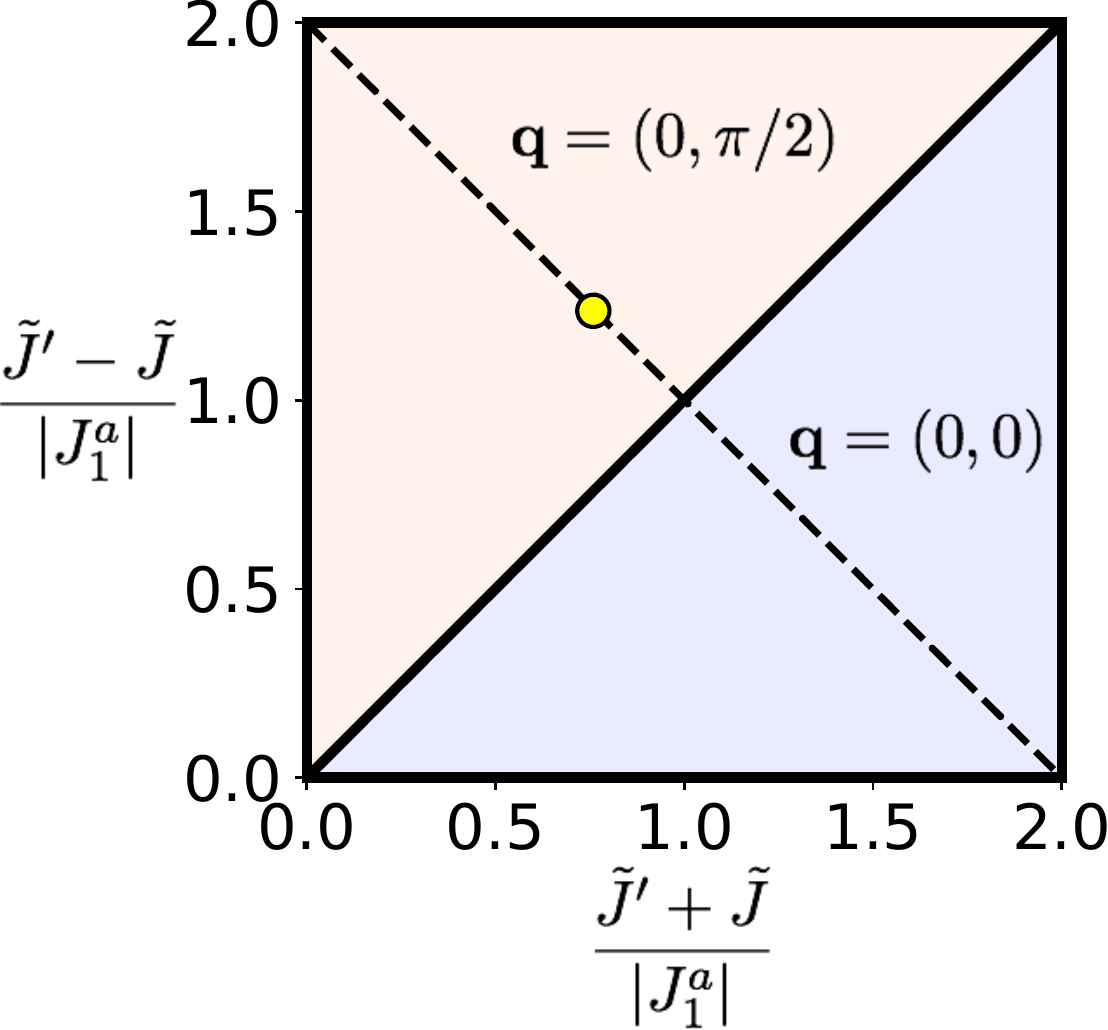}
\caption{\footnotesize{
Phase diagram showing the 1-magnon instability of the fully polarised phase.
For $\tilde{J}>0$ the dispersion goes soft at ${\bf q}=(0,0)$, implying a transition to 2-sublattice canted antiferromagnet.
For $\tilde{J}<0$ the dispersion goes soft at ${\bf q}=(0,\pi/2)$, implying a transition to 4-sublattice canted antiferromagnet.
At  $\tilde{J}=0$ there is a line of zeros in the dispersion (see Fig.~\ref{fig:dispersion}), which is a sign of high frustration.
The dashed line shows the path taken though configuration space in Fig.~\ref{fig:binding_energy}, and the yellow dot the parameters extracted for BaCdVO(PO$_4$)$_2$ in Ref.~\onlinecite{bhartiya19}.
}}
\label{fig:phase_diag}
\end{figure}

In particular, the location of the first 1-magnon instability of the fully polarised magnet is determined by the sign of $\tilde{J}$.
Assuming that $\tilde{J}^\prime > \tilde{J}$, it occurs at ${\bf q}=(0,0)$ for $\tilde{J}>0$ and at ${\bf q}=(0,\pi/2)$ for $\tilde{J}<0$ (the physics is just mirrored for $\tilde{J}^\prime < \tilde{J}$).
Thus a simple phase diagram can be plotted in terms of $(\tilde{J}^\prime + \tilde{J})/|J_1^a|$ and $(\tilde{J}^\prime - \tilde{J})/|J_1^a|$, as shown in Fig.~\ref{fig:phase_diag}.
The critical fields at which the instabilities take place are given by,
\begin{align}
h_{0} &= \tilde{J}^\prime +\tilde{J}, \quad {\bf q}=(0,0) \nonumber \\
h_{\pi/2} &= \tilde{J}^\prime, \quad {\bf q}=(0,\pi/2).
\end{align}
Since ${\bf q}$ is associated with a 2-site unit cell (see Fig.~\ref{fig:J_interactions}), an instability at ${\bf q}=(0,0)$ implies the development of 2-site magnetic ordering, while an instability at ${\bf q}=(0,\pi/2)$ implies 4-site magnetic ordering.

\begin{figure}[t]
\centering
\includegraphics[width=0.35\textwidth]{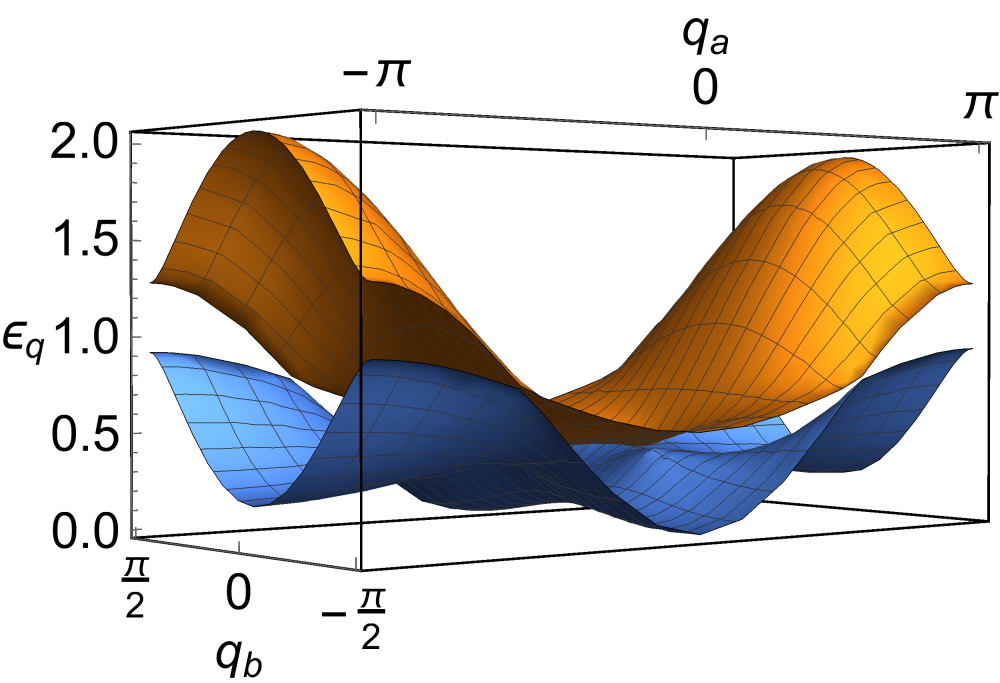}
\caption{\footnotesize{
Reduced dispersion relationship, $\epsilon_{\bf q}^{\alpha/\beta}$ [Eq.~\ref{eq:epsilon}], shown in the 2-site Brillouin zone for the parameters extracted in Ref.~\onlinecite{bhartiya19}.
The upper (yellow) branch, $\epsilon_{\bf q}^{\alpha}$, is gapped, while the lower (blue) band, $\epsilon_{\bf q}^{\beta}$, has a gapless point at ${\bf q}=(0,\pi/2)$.
The parameters are $J_1^a = J_1^{\prime a} = J_1^b = -0.42$ meV, $J_1^{\prime b} = -0.34$ meV, $J_2^+ = J_2^- = 0.16$ meV and $J_2^{\prime +} = J_2^{\prime -} = 0.38$ meV.
}}
\label{fig:dispersion_Bhartiya}
\end{figure}

Since BaCdVO(PO$_4$)$_2$ has 4-sublattice magnetic order at low-field \cite{skoulatos19,bhartiya19}, the main interest of this paper is in the instability at ${\bf q}=(0,\pi/2)$, and the associated chemical potential can be defined as $\mu = h_{\pi/2} -h$.
This can be used to define the shifted dispersion relations,
\begin{align}
 \epsilon_{\bf q}^{\alpha/\beta} =  \omega_{\bf q}^{\alpha/\beta} +\mu,
 \label{eq:epsilon}
\end{align}
such that the minima of $ \epsilon_{\bf q}^{\alpha/\beta}$ are fixed to zero energy, as can be seen in Fig.~\ref{fig:dispersion_Bhartiya}.

\begin{figure*}[t]
\centering
\includegraphics[width=0.9\textwidth]{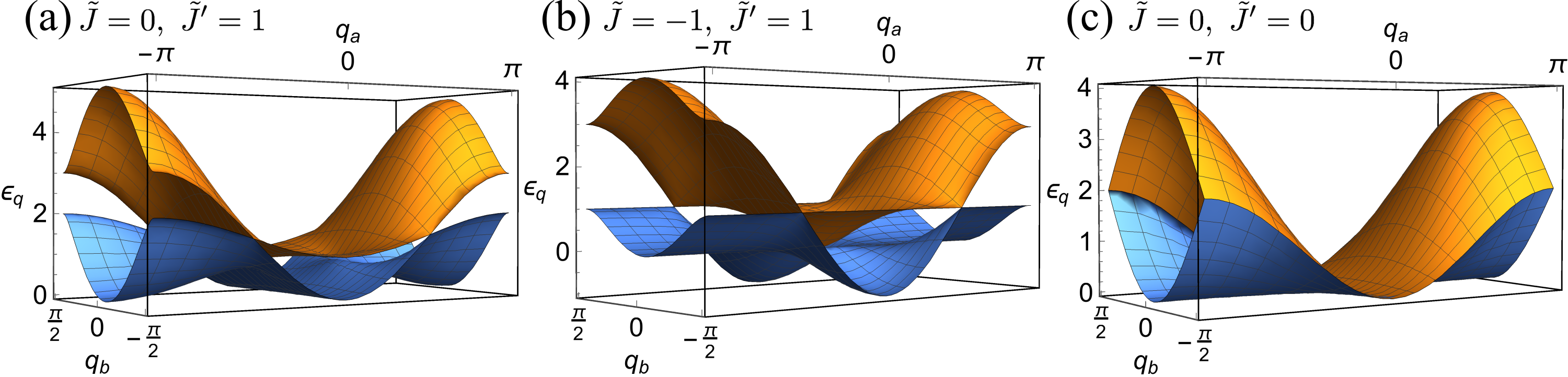}
\caption{\footnotesize{
The shifted dispersion of magnons in $\mathcal{H}_{\sf spin}$ [Eq.~\ref{eq:Hspin}].
The lower branch is $\epsilon_{\bf q}^{\beta}$ and the upper branch $ \epsilon_{\bf q}^{\alpha}$ [Eq.~\ref{eq:epsilon}].
(a) For $\tilde{J}=0$ there is a line of zeros running along $q_a=0$.
The plot is for $J_1^a=J_1^{\prime a}=J_1^b=J_1^{\prime b}=-1$, $J_2^+=J_2^-=0.5$ and  $J_2^{\prime +}=J_2^{\prime -}=1$.
(b) For $\tilde{J}<0$ there is a gapless point at ${\bf q}=(0,\pi/2)$. 
The plot is for $J_1^a=J_1^{\prime a}=J_1^b=J_1^{\prime b}=-1$, $J_2^+=J_2^-=0$ and  $J_2^{\prime +}=J_2^{\prime -}=1$.
For $\tilde{J}=\tilde{J}^\prime=0$ there are two lines of zeros, running along $q_a=0$ and $q_b=0$.
The plot is for $J_1^a=J_1^{\prime a}=J_1^b=J_1^{\prime b}=-1$, $J_2^+=J_2^-=-J_2^{\prime +}=J_2^{\prime -}=1$ (i.e. the highly frustrated point of the $J_1$-$J_2$ model).
}}
\label{fig:dispersion}
\end{figure*}

Certain points on the phase diagram shown in Fig.~\ref{fig:phase_diag} include the well-studied $J_1$-$J_2$ Heisenberg model.
Most interestingly $\tilde{J}=\tilde{J}^\prime=0$ includes the highly-frustrated point, $|J_1|= J_2/2$, where it has been shown that frustration destroys the possibility of forming long-range magnetic order, and instead a bond-nematic phase forms that persists all the way to $h=0$ \cite{shannon04,shannon06}.
The mechanism underpinning this is the partial localisation of the one-magnon excitations, which thus gain relatively little energy from hopping, and get outcompeted by more mobile bound-magnon pairs \cite{shannon06}.
This partial localisation can be seen in the 1-magnon dispersion as intersecting lines of zero energy modes in $ \epsilon_{\bf q}^{\beta}$, one along $q_a=0$ and the other along $q_b=0$ (see Fig.~\ref{fig:dispersion}).

This suggests that it is worth searching for other parameter sets with lines of zeros in $\epsilon_{\bf q}^{\alpha/\beta}$, since these are likely to correspond to parameters that are particularly favourable for the formation of bond-nematic order.
Inspection of the dispersion relations show that there is a line of parameters, defined by $\tilde{J}=0$ (i.e. the border between the ${\bf q}=(0,\pi/2)$ and ${\bf q}=(0,0)$ instabilities), for which the shifted dispersion has lines of zeros running along $q_a=0$, an example of which is shown in Fig.~\ref{fig:dispersion}.
Moving away from the $\tilde{J}=0$ line reduces the phase space of low-energy single-magnon modes (see Fig.~\ref{fig:dispersion}), and thus the expectation is that the magnon-binding energy should reduce.


\section{Magnon bound state}


The main focus of this paper is on the existence of magnon bound states in the fully-polarised phase of $\mathcal{H}_{\sf spin}$ [Eq.~\ref{eq:Hspin}] that condense preferentially as the magnetic field is lowered.
This results in $\langle S_i^-S_j^- \rangle$ taking a finite value, which is exactly the bond-nematic order parameter \cite{shannon06,ueda13}.

The calculation of magnon binding energies has been used to understand a number of model systems with competing ferromagnetic and antiferromagnetic interactions \cite{ueda09,ueda13,ueda15,smerald15}.
The novelty of the calculations presented here arises firstly from their application to the material BaCdVO(PO$_4$)$_2$ and secondly from the technical challenge posed by a 2-site unit cell.  

The basic idea is that, when decreasing magnetic field from a high value, bound magnon pairs condense before single-magnon excitations.
The single-magnon excitations condense when the dispersion minimum (i.e. the minimum of $\omega_{\bf q}^{\alpha/\beta}$ [Eq.~\ref{eq:H2ab}]) touches zero, resulting in a transition to a canted antiferromagnetic phase at $h=h_{\pi/2}$.
However, if a bound-magnon state exists below the 2-magnon continuum, then its dispersion minimum is at $-2\mu-\Delta$, with binding energy $\Delta>0$, resulting in condensation of bound magnon pairs at $h=h_{\pi/2}+\Delta/2$.
Thus the first instability of the fully-polarised state is not to a canted antiferromagnet, but instead to a partially-polarised bond nematic.

As the field is lowered further, the bond nematic may persist all the way to zero field, or it may transition into an ordered magnet, with the former case only occurring for very high frustration.
The lower bound for the width of the bond-nematic field range is given by translating $\Delta$ into field units, but the true value may be considerably larger.

Calculation of $\Delta$ requires a determination of the renormalised two-particle scattering vertex, starting from the bosonic Hamiltonian \cite{nakanishi69,ueda13,ueda15,smerald15}.
A bound state corresponds to a divergence of this renormalised vertex that occurs outside of the two-particle continuum, and thus results in a separate pole in the two-particle Green's function.
The renormalised two-particle vertex can be calculated exactly, allowing $\Delta$ to be determined without approximation.

The lowest energy 1-magnon excitations are the $\beta$ bosons, which have a dispersion minima at ${\bf q}=(0,\pm\pi/2)$.
Thus the lowest-energy bound state will be a pairing of $\beta$ bosons with total momentum ${\bf K}=0$.
As such the focus of the calculations will be on determining the renormalised vertex of the purely $\beta$ term in $\mathcal{H}_4$ [Eq.~\ref{eq:H4ab}]. 
In principle it is possible to also have bound pairs of $\alpha$ bosons or of one $\alpha$ and one $\beta$ boson, but, due to $\epsilon_{\bf q}^\alpha$ having a gap, these will not condense.
 
 \begin{figure*}[t]
\centering
\includegraphics[width=0.8\textwidth]{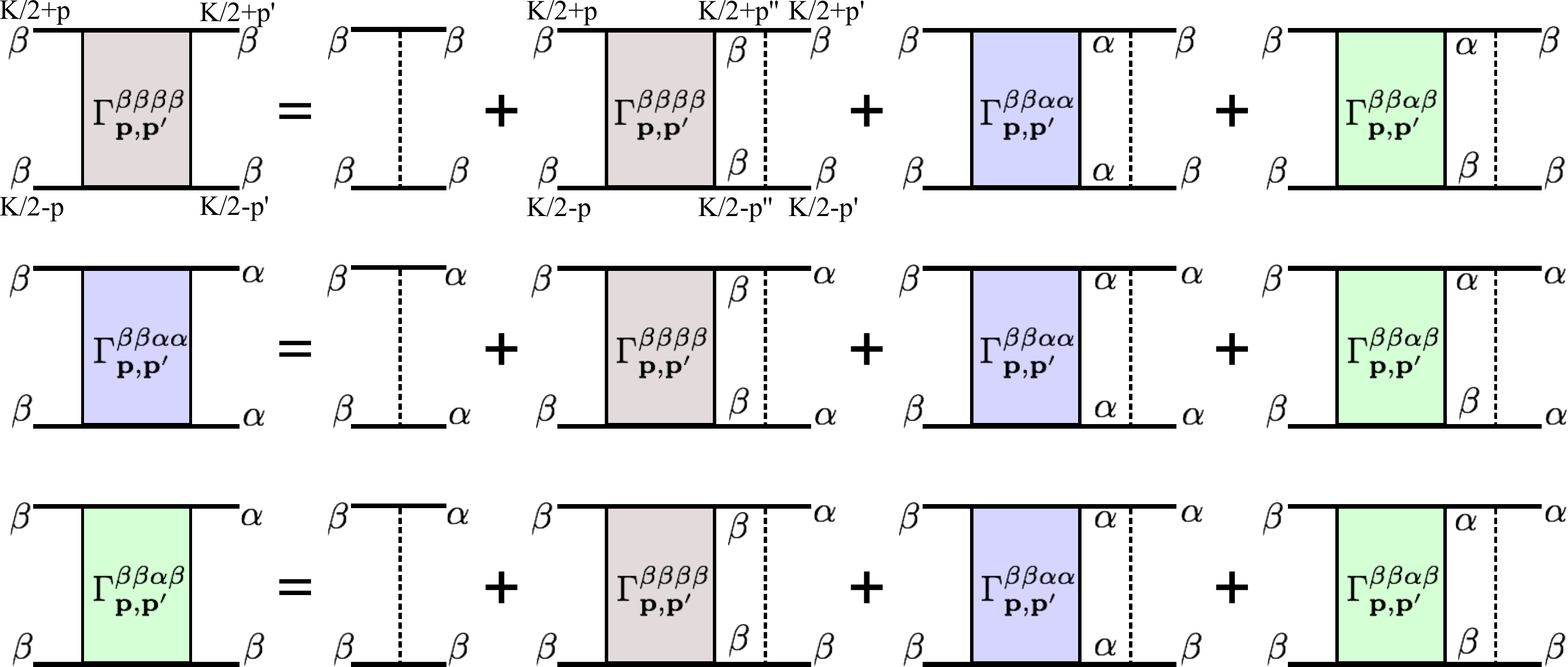}
\caption{\footnotesize{
Diagrammatic equations for the renomalised vertices $\Gamma^{\beta\beta\beta\beta}_{{\bf p},{\bf p}^{\prime}}$, $\Gamma^{\beta\beta\alpha\alpha}_{{\bf p},{\bf p}^{\prime}}$ and $\Gamma^{\beta\beta\alpha\beta}_{{\bf p},{\bf p}^{\prime}}$.
Dashed lines represent the bare $V$ vertices.
Divergence of one of these vertices indicates the existence of a magnon bound state.
}}
\label{fig:ladder_eq}
\end{figure*}

The renormalised vertex, $\Gamma_{{\bf p},{\bf p}^\prime}^{\beta\beta\beta\beta}$ can be calculated via the Bethe-Salpeter equation \cite{ueda13}, and, in the case of two flavours of bosons, it is necessary to consider a set of coupled equations.
These can be represented diagramatically, as shown in Fig.~\ref{fig:ladder_eq}.

The first problem to be overcome when solving these equations is that the bare interaction contains the infinite on-site potential $U$.
I again concentrate on the case $J_1^a=J_1^{\prime a}$, where eliminating $U$ (see Appendix~\ref{App:U}) results in a set of 6 simultaneous equations. 
The first two of these are,
\begin{align}
&\int \frac{d^2p^{\prime\prime}}{2\pi^2} \left[
\Gamma^{\beta\beta\beta\beta}_{{\bf p},{\bf p}^{\prime\prime}}g^{\beta\beta}_{{\bf p}^{\prime\prime}}
+\Gamma^{\beta\beta\alpha\alpha}_{{\bf p},{\bf p}^{\prime\prime}}g^{\alpha\alpha}_{{\bf p}^{\prime\prime}}
\right] =1 \nonumber\\
&\int \frac{d^2p^{\prime\prime}}{2\pi^2}
\Gamma^{\beta\beta\alpha\beta}_{{\bf p},{\bf p}^{\prime\prime}}g^{\alpha\beta}_{{\bf p}^{\prime\prime}}
=0,
\label{eq:row12}
\end{align}
where,
\begin{align}
g^{\gamma\delta}_{\bf p}({\bf K},\Delta) = \frac{1}{\epsilon^\gamma_{{\bf K}/2+{\bf p}^{\prime\prime}} + \epsilon^\delta_{{\bf K}/2-{\bf p}^{\prime\prime}} +\Delta - i0^+ },
\end{align}
with $\gamma\delta=\alpha\alpha,\alpha\beta,\beta\beta$.
The third is,
\begin{align}
& \langle \Gamma^{\beta\beta\beta\beta}_{{\bf p},{\bf p}^{\prime}} \rangle - \langle \Gamma^{\beta\beta\alpha\alpha}_{{\bf p},{\bf p}^{\prime}} \rangle =
\langle V^{\beta\beta\beta\beta}_{{\bf p},{\bf p}^{\prime}}\rangle - \langle V^{\beta\beta\alpha\alpha}_{{\bf p},{\bf p}^{\prime}}\rangle \nonumber \\
&\qquad - \int \frac{d^2p^{\prime\prime}}{2\pi^2} \left[ 
\Gamma^{\beta\beta\beta\beta}_{{\bf p},{\bf p}^{\prime\prime}} g^{\beta\beta}_{{\bf p}^{\prime\prime}} (\langle V^{\beta\beta\beta\beta}_{{\bf p}^{\prime\prime},{\bf p}^{\prime}}\rangle - \langle V^{\beta\beta\alpha\alpha}_{{\bf p}^{\prime\prime},{\bf p}^{\prime}}\rangle)  \right. \nonumber \\
&\qquad \qquad \qquad +\Gamma^{\beta\beta\alpha\alpha}_{{\bf p},{\bf p}^{\prime\prime}} g^{\alpha\alpha}_{{\bf p}^{\prime\prime}} (\langle V^{\alpha\alpha\beta\beta}_{{\bf p}^{\prime\prime},{\bf p}^{\prime}}\rangle -  \langle V^{\alpha\alpha\alpha\alpha}_{{\bf p}^{\prime\prime},{\bf p}^{\prime}}\rangle) \nonumber \\
&\qquad \qquad \qquad \left. +\Gamma^{\beta\beta\alpha\beta}_{{\bf p},{\bf p}^{\prime\prime}} g^{\alpha\beta}_{{\bf p}^{\prime\prime}} (\langle V^{\alpha\beta\beta\beta}_{{\bf p}^{\prime\prime},{\bf p}^{\prime}}\rangle -  \langle V^{\alpha\beta\alpha\alpha}_{{\bf p}^{\prime\prime},{\bf p}^{\prime}}\rangle)
\right],
\label{eq:row3}
\end{align}
where angular brackets are defined by $\langle \mathcal{O}_{{\bf p},{\bf p}^{\prime}} \rangle = \int \frac{d^2p^{\prime}}{2\pi^2} \mathcal{O}_{{\bf p},{\bf p}^{\prime}}$ and $V$ is the finite part of the vertex $\tilde{V}$.
The remaining 3 are,
\begin{align}
&\Gamma^{\beta\beta\gamma\delta}_{{\bf p},{\bf p}^{\prime}} = \langle \Gamma^{\beta\beta\gamma\delta}_{{\bf p},{\bf p}^{\prime}} \rangle 
+ \delta V^{\beta\beta\gamma\delta}_{{\bf p},{\bf p}^{\prime}} \nonumber \\
&- \int \frac{d^2p^{\prime\prime}}{2\pi^2} \left[ 
\Gamma^{\beta\beta\beta\beta}_{{\bf p},{\bf p}^{\prime\prime}} g^{\beta\beta}_{{\bf p}^{\prime\prime}} 
\delta V^{\beta\beta\gamma\delta}_{{\bf p}^{\prime\prime},{\bf p}^{\prime}} 
+\Gamma^{\beta\beta\alpha\alpha}_{{\bf p},{\bf p}^{\prime\prime}} g^{\alpha\alpha}_{{\bf p}^{\prime\prime}} 
\delta V^{\alpha\alpha\gamma\delta}_{{\bf p}^{\prime\prime},{\bf p}^{\prime}} \right. \nonumber \\
&\left. \qquad \qquad \qquad +\Gamma^{\beta\beta\alpha\beta}_{{\bf p},{\bf p}^{\prime\prime}} g^{\alpha\beta}_{{\bf p}^{\prime\prime}} 
\delta V^{\alpha\beta\gamma\delta}_{{\bf p}^{\prime\prime},{\bf p}^{\prime}} 
\right],
\label{eq:remaining_rows}
\end{align}
where,
\begin{align}
\delta V^{\beta\beta\beta\beta}_{{\bf p},{\bf p}^{\prime}}  = V^{\beta\beta\beta\beta}_{{\bf p},{\bf p}^{\prime}} - \langle V^{\beta\beta\beta\beta}_{{\bf p},{\bf p}^{\prime}} \rangle,
\end{align}
and similarly for the other $V$'s.

A good way to solve these equations is by expanding the renormalised vertices $\Gamma$, in terms of the same lattice harmonics as the bare vertices, $V$ \cite{ueda13}.
As a result, the $U$-independent simultaneous equations [i.e. Eq.~\ref{eq:row12}, Eq.~\ref{eq:row3} and Eq.~\ref{eq:remaining_rows}] can be used to define a 45 by 45 matrix (see Appendix~\ref{App:harmonics}).
When the determinant of the matrix is equal to zero, this corresponds to a divergence of the renormalised vertex $\Gamma^{\beta\beta\beta\beta}_{{\bf p},{\bf p}^{\prime}}$.
By calculating the value of $\Delta$ at which this occurs, the binding energy can be determined.
A useful check that no mistakes have been made in the construction of the matrix involves comparing to known results for the $J_1$-$J_2$ Heisenberg model on the square lattice \cite{smerald15}.
The same value of $\Delta$ was found from the more complicated calculation based on a 2-site unit cell described in this paper, as was obtained for the simpler 1-site unit-cell calculation that is possible for the $J_1$-$J_2$ model.
This lends confidence to the results at more general values of the $J$ parameters, where a 2-site unit cell is unavoidable.

The most interesting parameter set to investigate is the one determined for BaCdVO(PO$_4$)$_2$ from fits to inelastic neutron scattering experiments \cite{bhartiya19}.
For these parameters, which are listed under Eq.~\ref{eq:Hspin}, I find $\Delta = 0.032$ meV (equivalently $\Delta/|J_1^a| = 0.076$), implying that a bond nematic state forms below saturation.
Calculating the saturation field using $h_{\sf sat} = h_{\pi/2}+\Delta/2$ and $g_c=1.92$\cite{povarov19} gives $h_{\sf sat}= 3.9$T, whose low value compared to the experimentally determined $h_{\sf sat}\approx 6.5$T may indicate that the model parameters need to be slightly revised.
This wouldn't be too surprising, since it was reported in Ref.~\onlinecite{bhartiya19} that it was not possible to uniquely determine the model parameters from fits to the available experimental data.

One important point is that the 1-magnon spectrum, $\omega_{\bf q}$ [Eq.~\ref{eq:H2ab}], which is what is measured by inelastic neutron scattering, remains gapped at the saturation field, with a gap $h_{\sf sat}-h_{\pi/2} = \Delta/2$.
Measurement of a gap in the 1-magnon spectrum at the same field as thermodynamic measurements show the saturation transition to occur, would thus both provide evidence that the transition is not associated with a condensation of single magnons and allow an estimate of the magnon binding energy, $\Delta$.

\begin{figure}[t]
\centering
\includegraphics[width=0.35\textwidth]{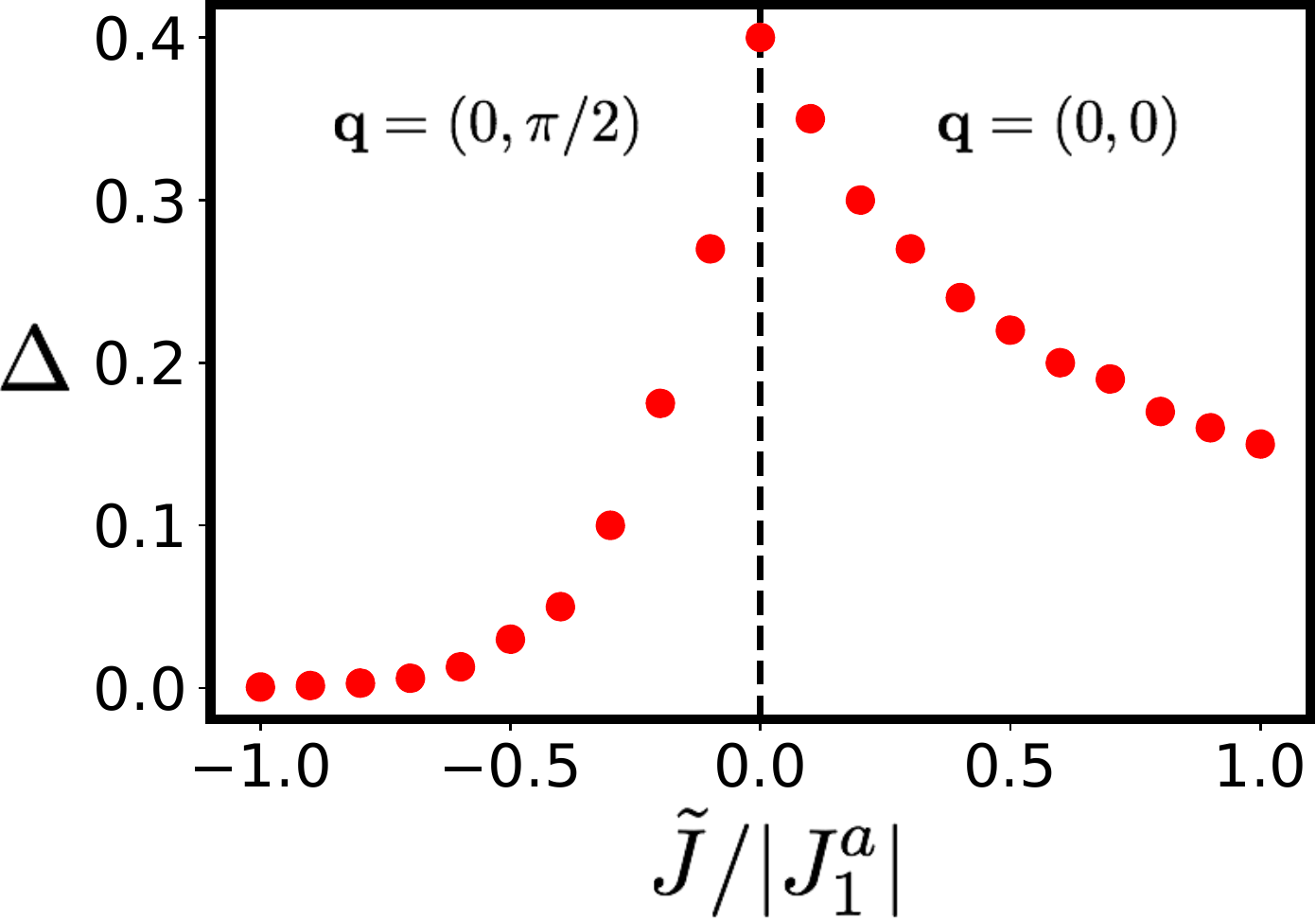}
\caption{\footnotesize{
Dependence of the magnon binding energy, $\Delta$, on the proximity to the highly frustrated line at $\tilde{J}=0$.
As expected, the magnon binding energy is enhanced by the build-up of low-energy modes close to $\tilde{J}=0$.
The parameters used are $J_1^a = J_1^{\prime a}  = J_1^b = J_1^{\prime b} =-1$, $0<J_2^+ = J_2^- <1$ and $J_2^{\prime +} = J_2^{\prime -} = 1$, corresponding to $-1 < \tilde{J}/|J_1^a|<1$. 
For $\tilde{J}<0$ the first 1-magnon instability is at ${\bf q}=(0,\pi/2)$ and for $\tilde{J}>0$ it is at ${\bf q}=(0,0)$ (see Fig.~\ref{fig:phase_diag}).
}}
\label{fig:binding_energy}
\end{figure}

Since further inelastic neutron scattering measurements may give a different parametrisation of the magnetic couplings in BaCdVO(PO$_4$)$_2$, and/or other oxyvanadate materials may be found to also be described by $\mathcal{H}_{\sf spin}$ [Eq.~\ref{eq:Hspin}], it is worth trying to explore parameter space to get a feeling for the robustness of the bond-nematic state.
While it would in principle be possible to explore fully the 8 dimensional space, it would be a thankless task.
Instead I consider a line of parameters with $\tilde{J}^\prime/|J_1^a| =1$ and $-1<\tilde{J}/|J_1^a| <1$  (see Fig.~\ref{fig:phase_diag}).
These parameters include both the region in which ${\bf q}=(0,\pi/2)$ is the first 1-magnon instability ($-1<\tilde{J}/|J_1^a| <0$) and in which ${\bf q}=(0,0)$ is the first 1-magnon instability ($0<\tilde{J}/|J_1^a| <1$).
As expected, the magnon binding energy, $\Delta$, is largest when $\tilde{J}=0$ (see Fig.~\ref{fig:binding_energy}), which I ascribe to the large phase space of low-energy modes associated with the line of zeros in $\epsilon_{\bf q}^{\beta}$ (see Fig.~\ref{fig:dispersion}).
Moving away from this point, $\Delta$ falls off quicker when moving in the $\tilde{J}<0$ (i.e. into the ${\bf q}=(0,\pi/2)$ region) than in the $\tilde{J}>0$ (i.e. into the ${\bf q}=(0,0)$ region), but in both directions remains positive in the full range $-1<\tilde{J}/|J_1^a| <1$.
This suggests that for ferromagnetic first-neighbour interactions and antiferromagnetic second-neighbour interactions the formation of a bond-nematic just below saturation is very robust.

The parameters for which $\Delta$ has been calculated in Fig.~\ref{fig:binding_energy} are $J_1^a=J_1^{\prime a}=J_1^b=J_1^{\prime b}=-1$, $0<J_2^+=J_2^-<1$ and  $J_2^{\prime +}=J_2^{\prime -}=1$.
The nice thing about these parameters is that they connect to the known value for the $J_1$-$J_2$ model for $J_2^+=J_2^-=1$ ($\Delta=0.148$) and that they can also be quantitatively compared to the maximally frustrated point of the $J_1$-$J_2$ model at $J_1=-1$ and $J_2=0.5$ \cite{smerald15}.
As expected, the value of $\Delta=0.52$ found at the maximally frustrated point is greater than the value of $\Delta=0.4$ at $J_2^+=J_2^-=0.5$ (i.e. on the highly-frustrated line).

Thus it seems possible to deduce two general principles, namely that the magnon binding energy is strongest when the parameters are close to the frustrated line defined by $\tilde{J}=0$, where there is a line of zeros in the shifted dispersion, and that this effect is enhanced approaching the highly-frustrated point familiar from the $J_1$-$J_2$ model, where two  lines of zeros intersect.
%


\section{Conclusion}


The main message of this paper is that BaCdVO(PO$_4$)$_2$ probably does realise a bond-nematic state at magnetic fields just below saturation.
More precisely, the magnetic Hamiltonian that is believed to describe BaCdVO(PO$_4$)$_2$ shows a robust tendency to bond-nematic ordering in the relevant parameter region.
This includes the set of parameters extracted from inelastic neutron scattering experiments \cite{bhartiya19}.
Further measurements may refine the magnetic-coupling parameters extracted in Ref.~\onlinecite{bhartiya19}, but, due to the robustness of magnon pairing to changes in model parameters, this is unlikely to change the conclusion that a bond-nematic state exists.

More generally, the robustness of the bond nematic state as the first instability of $\mathcal{H}_{\sf spin}$ [Eq.~\ref{eq:Hspin}] on lowering magnetic field, is related to the large phase space of low-energy modes in the 1-magnon dispersion.
This is most pronounced at certain lines in the phase diagram in which lines of zero-energy modes develop in the shifted dispersion relation.
%


\section*{Acknowledgments} 
I would like to thank Nic Shannon and Markos Skoulatos for many useful discussions, both about spin nematics in general and BaCdVO(PO$_4$)$_2$ in particular.
Thanks also to George Jackeli and Nic Shannon for useful comments on the manuscript.
I would also like to acknowledge many historical discussions with the late Hiroaki Ueda, who worked extensively on understanding magnon binding in frustrated ferromagnets.

\appendix


\section{The bosonic Hamiltonian for $J_1^a \neq J_1^{\prime a}$}
\label{App:bosonH}

In this Appendix a bosonic rewriting of $\mathcal{H}_{\sf spin}$ [Eq.~\ref{eq:Hspin}] is performed in the case of $J_1^a \neq J_1^{\prime a}$.

As already discussed in Section~\ref{Sec:Hamiltonian}, $\mathcal{H}_{\sf spin}$ can be exactly transformed into a Hamiltonian with 2 and 4 boson terms, as long as a hardcore constraint is enforced.
In the case of $J_1^a \neq J_1^{\prime a}$ the 2-boson part is given by,
\begin{align}
\mathcal{H}_2= \sum_{{\bf q}}
\left(  c\dg_{\bf q},d_{\bf q}\dg \right)
\left(
\begin{array}{cc}
A_{\bf q} &B_{\bf q}  \\
B_{\bf q}^* &A^\prime_{\bf q}   
\end{array}
\right) 
\left(
\begin{array}{c}
c_{\bf q}\pdg \\
d_{\bf q}\pdg
\end{array}
\right), 
\end{align}
where the difference from Eq.~\ref{eq:H2cd} is that the diagonal terms are no longer equal.
The new diagonal term is given by,
\begin{align}
A_{\bf q}^\prime &= h - J_1^{\prime a} (1-\cos q_a) - \frac{1}{2}(J_1^b + J_1^{\prime b}) \nonumber \\
&- \frac{1}{2}(J_2^+ + J_2^- + J_2^{\prime +} + J_2^{\prime -}),
\end{align}
while $A_{\bf q}$ and $B_{\bf q}$ are given in Eq.~\ref{eq:AB}.
As a result the eigenvectors of $\mathcal{H}_2$ are lengthier, and given by,
 \begin{align}
{\bf u}_{\bf q} &= \frac{1}{\mathcal{N}_u} \left(
\left[\frac{A_{\bf q}-A_{\bf q}^\prime}{2} + \sqrt{\left[\frac{A_{\bf q}-A_{\bf q}^\prime}{2} \right]^2 + |B_{\bf q}|^2 } \right]
\frac{B_{\bf q}}{|B_{\bf q}|^2},1
\right) \nonumber \\
{\bf v}_{\bf q} &= \frac{1}{\mathcal{N}_v} \left(
\left[\frac{A_{\bf q}-A_{\bf q}^\prime}{2} - \sqrt{\left[\frac{A_{\bf q}-A_{\bf q}^\prime}{2} \right]^2 + |B_{\bf q}|^2 } \right]
\frac{B_{\bf q}}{|B_{\bf q}|^2},1
\right)
\end{align}
with the normalisations,
 \begin{align}
\mathcal{N}_{u/v} &=  \frac{\sqrt{2}}{|B_{\bf q}|} 
\left[ |B_{\bf q}|^2+\left[\frac{A_{\bf q}-A_{\bf q}^\prime}{2} \right]^2 \right. \nonumber \\
&\left. \pm  \frac{A_{\bf q}-A_{\bf q}^\prime}{2} \sqrt{\left[\frac{A_{\bf q}-A_{\bf q}^\prime}{2} \right]^2 + |B_{\bf q}|^2 }\right]^{\frac{1}{2}}.
\end{align}
The resulting dispersion relations are,
 \begin{align}
 \omega_{\bf q}^{\alpha/\beta} = \frac{ A_{\bf q} + A_{\bf q}^\prime}{2}
 \pm \sqrt{  \left( \frac{ A_{\bf q} - A_{\bf q}^\prime}{2} \right)^2 +|B_{\bf q}|^2  }.
\end{align}

The 4-boson part of the Hamiltonian can be written as in Eq.~\ref{eq:Vcd}, but with,
 \begin{align}
&\tilde{V}^{cc}_{\bf q} = 2U+2J_1^a \cos q_a, \quad
V^{cd}_{\bf q} = 4B_{\bf q}, \nonumber \\
&\tilde{V}^{dd}_{\bf q} = 2U+2J_1^{\prime a} \cos q_a.
\end{align}
Transforming to the boson operators that diagonalise $\mathcal{H}_2$ gives,
\begin{align}
&\mathcal{H}_4 = \frac{1}{N} \sum_{{\bf k}_1...{\bf k}_4 } \left[
\tilde{V}^{\alpha\alpha\alpha\alpha}_{{\bf k}_1...{\bf k}_4}  \alpha\dg_{{\bf k}_1} \alpha\dg_{{\bf k}_2} \alpha_{{\bf k}_3} \alpha_{{\bf k}_4} \right. \nonumber \\
&+\tilde{V}^{\alpha\beta\alpha\alpha}_{{\bf k}_1...{\bf k}_4}  \alpha\dg_{{\bf k}_1} \beta\dg_{{\bf k}_2} \alpha_{{\bf k}_3} \alpha_{{\bf k}_4}  
+\tilde{V}^{\alpha\alpha\alpha\beta}_{{\bf k}_1...{\bf k}_4}  \alpha\dg_{{\bf k}_1} \alpha\dg_{{\bf k}_2} \alpha_{{\bf k}_3} \beta_{{\bf k}_4}  \nonumber \\
&+\tilde{V}^{\alpha\alpha\beta\beta}_{{\bf k}_1...{\bf k}_4}  \alpha\dg_{{\bf k}_1} \alpha\dg_{{\bf k}_2} \beta_{{\bf k}_3} \beta_{{\bf k}_4} 
+\tilde{V}^{\beta\beta\alpha\alpha}_{{\bf k}_1...{\bf k}_4}  \beta\dg_{{\bf k}_1} \beta\dg_{{\bf k}_2} \alpha_{{\bf k}_3} \alpha_{{\bf k}_4}  \nonumber \\
&+\tilde{V}^{\alpha\beta\alpha\beta}_{{\bf k}_1...{\bf k}_4}  \alpha\dg_{{\bf k}_1} \beta\dg_{{\bf k}_2} \alpha_{{\bf k}_3} \beta_{{\bf k}_4} 
+\tilde{V}^{\alpha\beta\beta\beta}_{{\bf k}_1...{\bf k}_4}  \alpha\dg_{{\bf k}_1} \beta\dg_{{\bf k}_2} \beta_{{\bf k}_3} \beta_{{\bf k}_4}  \nonumber \\
& \left. +\tilde{V}^{\beta\beta\alpha\beta}_{{\bf k}_1...{\bf k}_4}  \beta\dg_{{\bf k}_1} \beta\dg_{{\bf k}_2} \alpha_{{\bf k}_3} \beta_{{\bf k}_4}  + \tilde{V}^{\beta\beta\beta\beta}_{{\bf k}_1...{\bf k}_4}  \beta\dg_{{\bf k}_1} \beta\dg_{{\bf k}_2} \beta_{{\bf k}_3} \beta_{{\bf k}_4} \right] \nonumber \\
&\qquad \delta_{{\bf k}_1+{\bf k}_2-{\bf k}_3-{\bf k}_4}
\end{align}
where,
\begin{align}
\tilde{V}^{\alpha\alpha\alpha\alpha}_{{\bf k}_1...{\bf k}_4} &=  
u^{x*}_{{\bf k}_1} u^{x*}_{{\bf k}_2} u^x_{{\bf k}_3} u^x_{{\bf k}_4} \tilde{V}^{cc}_{{\bf k}_2-{\bf k}_4} 
+ u^{x*}_{{\bf k}_1} u^{y*}_{{\bf k}_2} u^x_{{\bf k}_3} u^y_{{\bf k}_4}  V^{cd}_{{\bf k}_2-{\bf k}_4} \nonumber \\
&+ u^{y*}_{{\bf k}_1} u^{y*}_{{\bf k}_2} u^y_{{\bf k}_3} u^y_{{\bf k}_4} \tilde{V}^{dd}_{{\bf k}_2-{\bf k}_4} \nonumber \\
\tilde{V}^{\alpha\beta\alpha\alpha}_{{\bf k}_1...{\bf k}_4} &=  
u^{x*}_{{\bf k}_1} v^{x*}_{{\bf k}_2} u^x_{{\bf k}_3} u^x_{{\bf k}_4} \tilde{V}^{cc}_{{\bf k}_2-{\bf k}_4} 
+ u^{x*}_{{\bf k}_1} v^{y*}_{{\bf k}_2} u^x_{{\bf k}_3} u^y_{{\bf k}_4}  V^{cd}_{{\bf k}_2-{\bf k}_4} \nonumber \\
&+ u^{y*}_{{\bf k}_1} v^{y*}_{{\bf k}_2} u^y_{{\bf k}_3} u^y_{{\bf k}_4} \tilde{V}^{dd}_{{\bf k}_2-{\bf k}_4} 
+ u^{x*}_{{\bf k}_1} v^{x*}_{{\bf k}_2} u^x_{{\bf k}_3} u^x_{{\bf k}_4} \tilde{V}^{cc}_{{\bf k}_1-{\bf k}_4} \nonumber \\
&+ u^{y*}_{{\bf k}_1} v^{x*}_{{\bf k}_2} u^x_{{\bf k}_3} u^y_{{\bf k}_4}  V^{cd}_{{\bf k}_1-{\bf k}_4} 
+ u^{y*}_{{\bf k}_1} v^{y*}_{{\bf k}_2} u^y_{{\bf k}_3} u^y_{{\bf k}_4} \tilde{V}^{dd}_{{\bf k}_1-{\bf k}_4} \nonumber \\
\tilde{V}^{\alpha\alpha\alpha\beta}_{{\bf k}_1...{\bf k}_4} &=  
u^{x*}_{{\bf k}_1} u^{x*}_{{\bf k}_2} u^x_{{\bf k}_3} v^x_{{\bf k}_4} \tilde{V}^{cc}_{{\bf k}_2-{\bf k}_4} 
+ u^{x*}_{{\bf k}_1} u^{y*}_{{\bf k}_2} u^x_{{\bf k}_3} v^y_{{\bf k}_4}  V^{cd}_{{\bf k}_2-{\bf k}_4} \nonumber \\
&+ u^{y*}_{{\bf k}_1} u^{y*}_{{\bf k}_2} u^y_{{\bf k}_3} v^y_{{\bf k}_4} \tilde{V}^{dd}_{{\bf k}_2-{\bf k}_4} 
+ u^{x*}_{{\bf k}_1} u^{x*}_{{\bf k}_2} u^x_{{\bf k}_3} v^x_{{\bf k}_4} \tilde{V}^{cc}_{{\bf k}_2-{\bf k}_3} \nonumber \\
&+ u^{x*}_{{\bf k}_1} u^{y*}_{{\bf k}_2} u^y_{{\bf k}_3} v^x_{{\bf k}_4}  V^{cd}_{{\bf k}_2-{\bf k}_3} 
+ u^{y*}_{{\bf k}_1} u^{y*}_{{\bf k}_2} u^y_{{\bf k}_3} v^y_{{\bf k}_4} \tilde{V}^{dd}_{{\bf k}_2-{\bf k}_3} \nonumber \\
\tilde{V}^{\alpha\alpha\beta\beta}_{{\bf k}_1...{\bf k}_4} &=  
u^{x*}_{{\bf k}_1} u^{x*}_{{\bf k}_2} v^x_{{\bf k}_3} v^x_{{\bf k}_4} \tilde{V}^{cc}_{{\bf k}_2-{\bf k}_4} 
+ u^{x*}_{{\bf k}_1} u^{y*}_{{\bf k}_2} v^x_{{\bf k}_3} v^y_{{\bf k}_4}  V^{cd}_{{\bf k}_2-{\bf k}_4} \nonumber \\
&+ u^{y*}_{{\bf k}_1} u^{y*}_{{\bf k}_2} v^y_{{\bf k}_3} v^y_{{\bf k}_4} \tilde{V}^{dd}_{{\bf k}_2-{\bf k}_4} \nonumber \\
\tilde{V}^{\beta\beta\alpha\alpha}_{{\bf k}_1...{\bf k}_4} &=  
v^{x*}_{{\bf k}_1} v^{x*}_{{\bf k}_2} u^x_{{\bf k}_3} u^x_{{\bf k}_4} \tilde{V}^{cc}_{{\bf k}_2-{\bf k}_4} 
+ v^{x*}_{{\bf k}_1} v^{y*}_{{\bf k}_2} u^x_{{\bf k}_3} u^y_{{\bf k}_4}  V^{cd}_{{\bf k}_2-{\bf k}_4} \nonumber \\
&+ v^{y*}_{{\bf k}_1} v^{y*}_{{\bf k}_2} u^y_{{\bf k}_3} u^y_{{\bf k}_4} \tilde{V}^{dd}_{{\bf k}_2-{\bf k}_4} \nonumber \\
\tilde{V}^{\alpha\beta\alpha\beta}_{{\bf k}_1...{\bf k}_4} &=  
u^{x*}_{{\bf k}_1} v^{x*}_{{\bf k}_2} u^x_{{\bf k}_3} v^x_{{\bf k}_4} \tilde{V}^{cc}_{{\bf k}_2-{\bf k}_4} 
+ u^{x*}_{{\bf k}_1} v^{y*}_{{\bf k}_2} u^x_{{\bf k}_3} v^y_{{\bf k}_4}  V^{cd}_{{\bf k}_2-{\bf k}_4} \nonumber \\
&+ u^{y*}_{{\bf k}_1} v^{y*}_{{\bf k}_2} u^y_{{\bf k}_3} v^y_{{\bf k}_4} \tilde{V}^{dd}_{{\bf k}_2-{\bf k}_4} 
+ u^{x*}_{{\bf k}_1} v^{x*}_{{\bf k}_2} u^x_{{\bf k}_3} v^x_{{\bf k}_4} \tilde{V}^{cc}_{{\bf k}_2-{\bf k}_3} \nonumber \\
&+ u^{x*}_{{\bf k}_1} v^{y*}_{{\bf k}_2} u^y_{{\bf k}_3} v^x_{{\bf k}_4}  V^{cd}_{{\bf k}_2-{\bf k}_3} 
+ u^{y*}_{{\bf k}_1} v^{y*}_{{\bf k}_2} u^y_{{\bf k}_3} v^y_{{\bf k}_4} \tilde{V}^{dd}_{{\bf k}_2-{\bf k}_3} \nonumber \\
&+ u^{x*}_{{\bf k}_1} v^{x*}_{{\bf k}_2} u^x_{{\bf k}_3} v^x_{{\bf k}_4} \tilde{V}^{cc}_{{\bf k}_1-{\bf k}_4} 
+ u^{y*}_{{\bf k}_1} v^{x*}_{{\bf k}_2} u^x_{{\bf k}_3} v^y_{{\bf k}_4}  V^{cd}_{{\bf k}_1-{\bf k}_4} \nonumber \\
&+ u^{y*}_{{\bf k}_1} v^{y*}_{{\bf k}_2} u^y_{{\bf k}_3} v^y_{{\bf k}_4} \tilde{V}^{dd}_{{\bf k}_1-{\bf k}_4} 
+ u^{x*}_{{\bf k}_1} v^{x*}_{{\bf k}_2} u^x_{{\bf k}_3} v^x_{{\bf k}_4} \tilde{V}^{cc}_{{\bf k}_1-{\bf k}_3} \nonumber \\ 
&+ u^{y*}_{{\bf k}_1} v^{x*}_{{\bf k}_2} u^y_{{\bf k}_3} v^x_{{\bf k}_4}  V^{cd}_{{\bf k}_1-{\bf k}_3} 
+ u^{y*}_{{\bf k}_1} v^{y*}_{{\bf k}_2} u^y_{{\bf k}_3} v^y_{{\bf k}_4} \tilde{V}^{dd}_{{\bf k}_1-{\bf k}_3} \nonumber \\
\tilde{V}^{\alpha\beta\beta\beta}_{{\bf k}_1...{\bf k}_4} &=  
u^{x*}_{{\bf k}_1} v^{x*}_{{\bf k}_2} v^x_{{\bf k}_3} v^x_{{\bf k}_4} \tilde{V}^{cc}_{{\bf k}_2-{\bf k}_4} 
+ u^{x*}_{{\bf k}_1} v^{y*}_{{\bf k}_2} v^x_{{\bf k}_3} v^y_{{\bf k}_4}  V^{cd}_{{\bf k}_2-{\bf k}_4} \nonumber \\
&+ u^{y*}_{{\bf k}_1} v^{y*}_{{\bf k}_2} v^y_{{\bf k}_3} v^y_{{\bf k}_4} \tilde{V}^{dd}_{{\bf k}_2-{\bf k}_4} 
+ u^{x*}_{{\bf k}_1} v^{x*}_{{\bf k}_2} v^x_{{\bf k}_3} v^x_{{\bf k}_4} \tilde{V}^{cc}_{{\bf k}_1-{\bf k}_4} \nonumber \\
&+ u^{y*}_{{\bf k}_1} v^{x*}_{{\bf k}_2} v^x_{{\bf k}_3} v^y_{{\bf k}_4}  V^{cd}_{{\bf k}_1-{\bf k}_4} 
+ u^{y*}_{{\bf k}_1} v^{y*}_{{\bf k}_2} v^y_{{\bf k}_3} v^y_{{\bf k}_4} \tilde{V}^{dd}_{{\bf k}_1-{\bf k}_4} \nonumber \\
\tilde{V}^{\beta\beta\alpha\beta}_{{\bf k}_1...{\bf k}_4} &=  
v^{x*}_{{\bf k}_1} v^{x*}_{{\bf k}_2} u^x_{{\bf k}_3} v^x_{{\bf k}_4} \tilde{V}^{cc}_{{\bf k}_2-{\bf k}_4} 
+ v^{x*}_{{\bf k}_1} v^{y*}_{{\bf k}_2} u^x_{{\bf k}_3} v^y_{{\bf k}_4}  V^{cd}_{{\bf k}_2-{\bf k}_4} \nonumber \\
&+ v^{y*}_{{\bf k}_1} v^{y*}_{{\bf k}_2} u^y_{{\bf k}_3} v^y_{{\bf k}_4} \tilde{V}^{dd}_{{\bf k}_2-{\bf k}_4} 
+ v^{x*}_{{\bf k}_1} v^{x*}_{{\bf k}_2} u^x_{{\bf k}_3} v^x_{{\bf k}_4} \tilde{V}^{cc}_{{\bf k}_2-{\bf k}_3} \nonumber \\ 
&+ v^{x*}_{{\bf k}_1} v^{y*}_{{\bf k}_2} u^y_{{\bf k}_3} v^x_{{\bf k}_4}  V^{cd}_{{\bf k}_2-{\bf k}_3} 
+ v^{y*}_{{\bf k}_1} v^{y*}_{{\bf k}_2} u^y_{{\bf k}_3} v^y_{{\bf k}_4} \tilde{V}^{dd}_{{\bf k}_2-{\bf k}_3} \nonumber \\
\tilde{V}^{\beta\beta\beta\beta}_{{\bf k}_1...{\bf k}_4} &=
v^{x*}_{{\bf k}_1} v^{x*}_{{\bf k}_2} v^x_{{\bf k}_3} v^x_{{\bf k}_4} \tilde{V}^{cc}_{{\bf k}_2-{\bf k}_4} 
+ v^{x*}_{{\bf k}_1} v^{y*}_{{\bf k}_2} v^x_{{\bf k}_3} v^y_{{\bf k}_4}  V^{cd}_{{\bf k}_2-{\bf k}_4} \nonumber \\
&+ v^{y*}_{{\bf k}_1} v^{y*}_{{\bf k}_2} v^y_{{\bf k}_3} v^y_{{\bf k}_4} \tilde{V}^{dd}_{{\bf k}_2-{\bf k}_4} . \nonumber \\
\label{eq:V}
\end{align}


\section{Eliminating the infinite potential $U$}
\label{App:U}

In order to solve the coupled ladder equations shown diagramatically in Fig.~\ref{fig:ladder_eq}, it is necessary to first eliminate the infinite potential $U$ \cite{nikuni95,jackeli04}.

The bare interaction vertices [Eq.~\ref{eq:V}] can always be rewritten so as to separate out the infinite part as,
\begin{align}
\tilde{V}^{\alpha\alpha\alpha\alpha}_{{\bf p},{\bf p}^\prime}  = 
V^{\alpha\alpha\alpha\alpha}_{{\bf p},{\bf p}^\prime} + 2U W^{\alpha\alpha\alpha\alpha}_{{\bf p},{\bf p}^\prime},
\label{eq:V=V+UW}
\end{align}
and similarly for the other $V$'s, where the ${\bf K}$ dependence has been suppressed for brevity.
In consequence the three equations shown in Fig.~\ref{fig:ladder_eq} can be written as,
\begin{widetext}
\begin{align}
 \Gamma^{\beta\beta\beta\beta}_{{\bf p},{\bf p}^{\prime}}  &= 
 V^{\beta\beta\beta\beta}_{{\bf p},{\bf p}^{\prime}}
- \int \frac{d^2p^{\prime\prime}}{2\pi^2} \left[ 
\Gamma^{\beta\beta\beta\beta}_{{\bf p},{\bf p}^{\prime\prime}} g^{\beta\beta}_{{\bf p}^{\prime\prime}}  V^{\beta\beta\beta\beta}_{{\bf p}^{\prime\prime},{\bf p}^{\prime}}
+\Gamma^{\beta\beta\alpha\alpha}_{{\bf p},{\bf p}^{\prime\prime}} g^{\alpha\alpha}_{{\bf p}^{\prime\prime}}  V^{\alpha\alpha\beta\beta}_{{\bf p}^{\prime\prime},{\bf p}^{\prime}}
+\Gamma^{\beta\beta\alpha\beta}_{{\bf p},{\bf p}^{\prime\prime}} g^{\alpha\beta}_{{\bf p}^{\prime\prime}}  V^{\alpha\beta\beta\beta}_{{\bf p}^{\prime\prime},{\bf p}^{\prime}}
\right] \nonumber \\
&+2U \left[  W^{\beta\beta\beta\beta}_{{\bf p},{\bf p}^{\prime}}
- \int \frac{d^2p^{\prime\prime}}{2\pi^2} \left[ 
\Gamma^{\beta\beta\beta\beta}_{{\bf p},{\bf p}^{\prime\prime}} g^{\beta\beta}_{{\bf p}^{\prime\prime}}  W^{\beta\beta\beta\beta}_{{\bf p}^{\prime\prime},{\bf p}^{\prime}}
+\Gamma^{\beta\beta\alpha\alpha}_{{\bf p},{\bf p}^{\prime\prime}} g^{\alpha\alpha}_{{\bf p}^{\prime\prime}}  W^{\alpha\alpha\beta\beta}_{{\bf p}^{\prime\prime},{\bf p}^{\prime}}
+\Gamma^{\beta\beta\alpha\beta}_{{\bf p},{\bf p}^{\prime\prime}} g^{\alpha\beta}_{{\bf p}^{\prime\prime}}  W^{\alpha\beta\beta\beta}_{{\bf p}^{\prime\prime},{\bf p}^{\prime}}
\right]  \right] \nonumber \\
 \Gamma^{\beta\beta\alpha\alpha}_{{\bf p},{\bf p}^{\prime}}  &= 
 V^{\beta\beta\alpha\alpha}_{{\bf p},{\bf p}^{\prime}}
- \int \frac{d^2p^{\prime\prime}}{2\pi^2} \left[ 
\Gamma^{\beta\beta\beta\beta}_{{\bf p},{\bf p}^{\prime\prime}} g^{\beta\beta}_{{\bf p}^{\prime\prime}}  V^{\beta\beta\alpha\alpha}_{{\bf p}^{\prime\prime},{\bf p}^{\prime}}
+\Gamma^{\beta\beta\alpha\alpha}_{{\bf p},{\bf p}^{\prime\prime}} g^{\alpha\alpha}_{{\bf p}^{\prime\prime}}  V^{\alpha\alpha\alpha\alpha}_{{\bf p}^{\prime\prime},{\bf p}^{\prime}}
+\Gamma^{\beta\beta\alpha\beta}_{{\bf p},{\bf p}^{\prime\prime}} g^{\alpha\beta}_{{\bf p}^{\prime\prime}}  V^{\alpha\beta\alpha\alpha}_{{\bf p}^{\prime\prime},{\bf p}^{\prime}}
\right] \nonumber \\
&+2U \left[  W^{\beta\beta\alpha\alpha}_{{\bf p},{\bf p}^{\prime}} 
- \int \frac{d^2p^{\prime\prime}}{2\pi^2} \left[ 
\Gamma^{\beta\beta\beta\beta}_{{\bf p},{\bf p}^{\prime\prime}} g^{\beta\beta}_{{\bf p}^{\prime\prime}}  W^{\beta\beta\alpha\alpha}_{{\bf p}^{\prime\prime},{\bf p}^{\prime}}
+\Gamma^{\beta\beta\alpha\alpha}_{{\bf p},{\bf p}^{\prime\prime}} g^{\alpha\alpha}_{{\bf p}^{\prime\prime}}  W^{\alpha\alpha\alpha\alpha}_{{\bf p}^{\prime\prime},{\bf p}^{\prime}}
+\Gamma^{\beta\beta\alpha\beta}_{{\bf p},{\bf p}^{\prime\prime}} g^{\alpha\beta}_{{\bf p}^{\prime\prime}}  W^{\alpha\beta\alpha\alpha}_{{\bf p}^{\prime\prime},{\bf p}^{\prime}}
\right]  \right] \nonumber \\
 \Gamma^{\beta\beta\alpha\beta}_{{\bf p},{\bf p}^{\prime}}  &= 
 V^{\beta\beta\alpha\beta}_{{\bf p},{\bf p}^{\prime}}
- \int \frac{d^2p^{\prime\prime}}{2\pi^2} \left[ 
\Gamma^{\beta\beta\beta\beta}_{{\bf p},{\bf p}^{\prime\prime}} g^{\beta\beta}_{{\bf p}^{\prime\prime}}  V^{\beta\beta\alpha\beta}_{{\bf p}^{\prime\prime},{\bf p}^{\prime}}
+\Gamma^{\beta\beta\alpha\alpha}_{{\bf p},{\bf p}^{\prime\prime}} g^{\alpha\alpha}_{{\bf p}^{\prime\prime}}  V^{\alpha\alpha\alpha\beta}_{{\bf p}^{\prime\prime},{\bf p}^{\prime}}
+\Gamma^{\beta\beta\alpha\beta}_{{\bf p},{\bf p}^{\prime\prime}} g^{\alpha\beta}_{{\bf p}^{\prime\prime}}  V^{\alpha\beta\alpha\beta}_{{\bf p}^{\prime\prime},{\bf p}^{\prime}}
\right] \nonumber \\
&+2U \left[  W^{\beta\beta\alpha\beta}_{{\bf p},{\bf p}^{\prime}} 
- \int \frac{d^2p^{\prime\prime}}{2\pi^2} \left[ 
\Gamma^{\beta\beta\beta\beta}_{{\bf p},{\bf p}^{\prime\prime}} g^{\beta\beta}_{{\bf p}^{\prime\prime}}  W^{\beta\beta\alpha\beta}_{{\bf p}^{\prime\prime},{\bf p}^{\prime}}
+\Gamma^{\beta\beta\alpha\alpha}_{{\bf p},{\bf p}^{\prime\prime}} g^{\alpha\alpha}_{{\bf p}^{\prime\prime}}  W^{\alpha\alpha\alpha\beta}_{{\bf p}^{\prime\prime},{\bf p}^{\prime}}
+\Gamma^{\beta\beta\alpha\beta}_{{\bf p},{\bf p}^{\prime\prime}} g^{\alpha\beta}_{{\bf p}^{\prime\prime}}  W^{\alpha\beta\alpha\beta}_{{\bf p}^{\prime\prime},{\bf p}^{\prime}}
\right]  \right]. \nonumber \\
\end{align}
\end{widetext}

In the case of $J_1^a=J_1^{\prime a}$ the $W$'s simplify to,
\begin{align}
&W^{\alpha\alpha\alpha\alpha}_{{\bf p},{\bf p}^\prime} 
=W^{\alpha\alpha\beta\beta}_{{\bf p},{\bf p}^\prime} 
=W^{\beta\beta\alpha\alpha}_{{\bf p},{\bf p}^\prime} 
=W^{\beta\beta\beta\beta}_{{\bf p},{\bf p}^\prime} = 1/2 \nonumber \\
&W^{\alpha\alpha\alpha\beta}_{{\bf p},{\bf p}^\prime} 
=W^{\alpha\beta\alpha\alpha}_{{\bf p},{\bf p}^\prime} 
=W^{\alpha\beta\beta\beta}_{{\bf p},{\bf p}^\prime} 
=W^{\beta\alpha\beta\beta}_{{\bf p},{\bf p}^\prime} =0 \nonumber \\
& W^{\alpha\beta\alpha\beta}_{{\bf p},{\bf p}^\prime} =2,
\end{align}
resulting in,
\begin{widetext}
\begin{align}
 \Gamma^{\beta\beta\beta\beta}_{{\bf p},{\bf p}^{\prime}}  &= 
 V^{\beta\beta\beta\beta}_{{\bf p},{\bf p}^{\prime}}
- \int \frac{d^2p^{\prime\prime}}{2\pi^2} \left[ 
\Gamma^{\beta\beta\beta\beta}_{{\bf p},{\bf p}^{\prime\prime}} g^{\beta\beta}_{{\bf p}^{\prime\prime}}  V^{\beta\beta\beta\beta}_{{\bf p}^{\prime\prime},{\bf p}^{\prime}}
+\Gamma^{\beta\beta\alpha\alpha}_{{\bf p},{\bf p}^{\prime\prime}} g^{\alpha\alpha}_{{\bf p}^{\prime\prime}}  V^{\alpha\alpha\beta\beta}_{{\bf p}^{\prime\prime},{\bf p}^{\prime}}
+\Gamma^{\beta\beta\alpha\beta}_{{\bf p},{\bf p}^{\prime\prime}} g^{\alpha\beta}_{{\bf p}^{\prime\prime}}  V^{\alpha\beta\beta\beta}_{{\bf p}^{\prime\prime},{\bf p}^{\prime}}
\right] \nonumber \\
&+U \left[  1
- \int \frac{d^2p^{\prime\prime}}{2\pi^2} \left[ 
\Gamma^{\beta\beta\beta\beta}_{{\bf p},{\bf p}^{\prime\prime}} g^{\beta\beta}_{{\bf p}^{\prime\prime}}  
+\Gamma^{\beta\beta\alpha\alpha}_{{\bf p},{\bf p}^{\prime\prime}} g^{\alpha\alpha}_{{\bf p}^{\prime\prime}}  
\right]  \right] \nonumber \\
 \Gamma^{\beta\beta\alpha\alpha}_{{\bf p},{\bf p}^{\prime}}  &= 
 V^{\beta\beta\alpha\alpha}_{{\bf p},{\bf p}^{\prime}}
- \int \frac{d^2p^{\prime\prime}}{2\pi^2} \left[ 
\Gamma^{\beta\beta\beta\beta}_{{\bf p},{\bf p}^{\prime\prime}} g^{\beta\beta}_{{\bf p}^{\prime\prime}}  V^{\beta\beta\alpha\alpha}_{{\bf p}^{\prime\prime},{\bf p}^{\prime}}
+\Gamma^{\beta\beta\alpha\alpha}_{{\bf p},{\bf p}^{\prime\prime}} g^{\alpha\alpha}_{{\bf p}^{\prime\prime}}  V^{\alpha\alpha\alpha\alpha}_{{\bf p}^{\prime\prime},{\bf p}^{\prime}}
+\Gamma^{\beta\beta\alpha\beta}_{{\bf p},{\bf p}^{\prime\prime}} g^{\alpha\beta}_{{\bf p}^{\prime\prime}}  V^{\alpha\beta\alpha\alpha}_{{\bf p}^{\prime\prime},{\bf p}^{\prime}}
\right] \nonumber \\
&+U \left[  1
- \int \frac{d^2p^{\prime\prime}}{2\pi^2} \left[ 
\Gamma^{\beta\beta\beta\beta}_{{\bf p},{\bf p}^{\prime\prime}} g^{\beta\beta}_{{\bf p}^{\prime\prime}}  
+\Gamma^{\beta\beta\alpha\alpha}_{{\bf p},{\bf p}^{\prime\prime}} g^{\alpha\alpha}_{{\bf p}^{\prime\prime}}  
\right]  \right] \nonumber \\
 \Gamma^{\beta\beta\alpha\beta}_{{\bf p},{\bf p}^{\prime}}  &= 
 V^{\beta\beta\alpha\beta}_{{\bf p},{\bf p}^{\prime}}
- \int \frac{d^2p^{\prime\prime}}{2\pi^2} \left[ 
\Gamma^{\beta\beta\beta\beta}_{{\bf p},{\bf p}^{\prime\prime}} g^{\beta\beta}_{{\bf p}^{\prime\prime}}  V^{\beta\beta\alpha\beta}_{{\bf p}^{\prime\prime},{\bf p}^{\prime}}
+\Gamma^{\beta\beta\alpha\alpha}_{{\bf p},{\bf p}^{\prime\prime}} g^{\alpha\alpha}_{{\bf p}^{\prime\prime}}  V^{\alpha\alpha\alpha\beta}_{{\bf p}^{\prime\prime},{\bf p}^{\prime}}
+\Gamma^{\beta\beta\alpha\beta}_{{\bf p},{\bf p}^{\prime\prime}} g^{\alpha\beta}_{{\bf p}^{\prime\prime}}  V^{\alpha\beta\alpha\beta}_{{\bf p}^{\prime\prime},{\bf p}^{\prime}}
\right] \nonumber \\
&-4U 
\int \frac{d^2p^{\prime\prime}}{2\pi^2} 
\Gamma^{\beta\beta\alpha\beta}_{{\bf p},{\bf p}^{\prime\prime}} g^{\alpha\beta}_{{\bf p}^{\prime\prime}}  .
  \nonumber \\
\end{align}
\end{widetext}
From these equations one can derive Eqs.~\ref{eq:row12}, Eq.~\ref{eq:row3} and Eqs.~\ref{eq:remaining_rows} in the main text.
Eqs.~\ref{eq:row12} come from setting each of the coefficients of $U$ to be 0, which is necessary to ensure that the vertex functions are finite for arbitrary parameter values.
Eq.~\ref{eq:row3} is derived by taking the difference between the two first equations above, and then applying the average $\langle \mathcal{O}_{{\bf p},{\bf p}^{\prime}} \rangle = \int \frac{d^2p^{\prime}}{2\pi^2} \mathcal{O}_{{\bf p},{\bf p}^{\prime}}$.
Finally Eqs.~\ref{eq:remaining_rows} are derived by taking $\Gamma - \langle \Gamma \rangle$, and thus eliminating the infinite part, which does not depend on ${\bf p}^\prime$.


\section{Expansion of Bethe-Salpeter equations in lattice harmonics}
\label{App:harmonics}

One way to solve the $U$-independent simultaneous equations, Eq.~\ref{eq:row12}, Eq.~\ref{eq:row3} and Eq.~\ref{eq:remaining_rows}, is to expand both the $V$'s and $\Gamma$'s in the same set of lattice harmonics.
The natural choise is the lattice harmonics of the $V$'s, which can be grouped into a vector,
\begin{align}
&{\bf T}({\bf p}) = 
\left( 
1,e^{ip_a},e^{-ip_a}, 
\frac{B_{\bf p}}{|B_{\bf p}|} e^{ip_b}, \frac{B_{\bf p}}{|B_{\bf p}|} e^{-ip_b}, \right. \nonumber \\
&\frac{B_{-{\bf p}}}{|B_{\bf p}|} e^{-ip_b}, \frac{B_{-{\bf p}}}{|B_{\bf p}|} e^{ip_b},
 \frac{B_{\bf p}}{|B_{\bf p}|} e^{i(pa+pb)}, \frac{B_{\bf p}}{|B_{\bf p}|} e^{-i(pa+pb)}, \nonumber \\
& \frac{B_{-{\bf p}}}{|B_{\bf p}|} e^{-i(pa+pb)}, \frac{B_{-{\bf p}}}{|B_{\bf p}|} e^{i(pa+pb)},  \frac{B_{\bf p}}{|B_{\bf p}|} e^{i(pa-pb)}, \nonumber \\
&\left.  \frac{B_{\bf p}}{|B_{\bf p}|} e^{-i(pa-pb)}, \frac{B_{-{\bf p}}}{|B_{\bf p}|} e^{-i(pa-pb)}, \frac{B_{-{\bf p}}}{|B_{\bf p}|} e^{i(pa-pb)} 
\right),
\end{align}
where $B_{\bf p}$ is given in Eq.~\ref{eq:AB}.
In terms of this vector the $V$'s can be written as,
\begin{align}
V_{{\bf p},{\bf p}^\prime} = {\bf T}^*({\bf p}) \cdot  {\bf J} \cdot {\bf T}({\bf p}^\prime),
\end{align}
where ${\bf J}$ is a 15 by 15 matrix.
This separates the ${\bf p}$ and ${\bf p}^\prime$ dependence of the $V$'s.
The matrix is block diagonal, and can be written as,
\begin{align}
{\bf J} =
\left(
\begin{array}{ccccc}
0 & 0 & 0 & 0 & 0    \\
0 & {\bf M}_a & 0 & 0 & 0   \\
0 & 0 & {\bf M}_b & 0 & 0   \\
0 & 0 & 0 & {\bf M}_- & 0   \\
0 & 0 & 0 & 0 & {\bf M}_+   
\end{array}
\right),
\end{align}
where ${\bf M}_a$ is a 2 by 2 matrix and ${\bf M}_b$, ${\bf M}_-$ and ${\bf M}_+$ are all 4 by 4 matrices.

Starting from Eq.~\ref{eq:V}, each of the $V$'s can be considered in turn.
The result is $V_{{\bf p},{\bf p}^\prime}^{\alpha\alpha\alpha\alpha} =  V_{{\bf p},{\bf p}^\prime}^{\beta\beta\beta\beta}$, with ${\bf J}^{\alpha\alpha\alpha\alpha}$ given by,
\begin{align}
{\bf M}_a^{\alpha\alpha\alpha\alpha} &=
\frac{1}{2}\left(
\begin{array}{cc}
J_1^a & 0    \\
0 & J_1^a   
\end{array}
\right), \nonumber \\
{\bf M}_b^{\alpha\alpha\alpha\alpha} &=
\frac{1}{2}\left(
\begin{array}{cccc}
J_1^b & 0 & 0 & 0   \\
0 & J_1^{\prime b} & 0 & 0   \\
0 & 0 & 0 & 0   \\
0 & 0 & 0 & 0   \\  
\end{array}
\right), \nonumber \\
{\bf M}_-^{\alpha\alpha\alpha\alpha} &=
\frac{1}{2}\left(
\begin{array}{cccc}
J_2^- & 0 & 0 & 0   \\
0 & J_2^{\prime -} & 0 & 0   \\
0 & 0 & 0 & 0   \\
0 & 0 & 0 & 0   \\  
\end{array}
\right), \nonumber \\
{\bf M}_+^{\alpha\alpha\alpha\alpha} &=
\frac{1}{2}\left(
\begin{array}{cccc}
J_2^{\prime +} & 0 & 0 & 0   \\
0 & J_2^+ & 0 & 0   \\
0 & 0 & 0 & 0   \\
0 & 0 & 0 & 0   \\  
\end{array}
\right),
\end{align}
$V_{{\bf p},{\bf p}^\prime}^{\alpha\alpha\alpha\beta} =  - V_{{\bf p},{\bf p}^\prime}^{\beta\beta\alpha\beta}$, with ${\bf J}^{\alpha\alpha\alpha\beta}$ given by,
\begin{align}
{\bf M}_a^{\alpha\alpha\alpha\beta} &=
\frac{1}{2}\left(
\begin{array}{cc}
0 & 0    \\
0 & 0   
\end{array}
\right), \nonumber \\
{\bf M}_b^{\alpha\alpha\alpha\beta} &=
\frac{1}{2}\left(
\begin{array}{cccc}
J_1^b & 0 & -J_1^b  & 0   \\
0 & J_1^{\prime b} & 0 & -J_1^{\prime b}   \\
0 & 0 & 0 & 0   \\
0 & 0 & 0 & 0   \\  
\end{array}
\right), \nonumber \\
{\bf M}_-^{\alpha\alpha\alpha\beta} &=
\frac{1}{2}\left(
\begin{array}{cccc}
J_2^- & 0 & -J_2^- & 0   \\
0 & J_2^{\prime -} & 0 & -J_2^{\prime -}   \\
0 & 0 & 0 & 0   \\
0 & 0 & 0 & 0   \\  
\end{array}
\right), \nonumber \\
{\bf M}_+^{\alpha\alpha\alpha\beta} &=
\frac{1}{2}\left(
\begin{array}{cccc}
J_2^{\prime +} & 0 & -J_2^{\prime +} & 0   \\
0 & J_2^+ & 0 & -J_2^+   \\
0 & 0 & 0 & 0   \\
0 & 0 & 0 & 0   \\  
\end{array}
\right),
\end{align}
$V_{{\bf p},{\bf p}^\prime}^{\alpha\beta\alpha\alpha} =  - V_{{\bf p},{\bf p}^\prime}^{\alpha\beta\beta\beta}$, with ${\bf J}^{\alpha\beta\alpha\alpha} = ({\bf J}^{\alpha\alpha\alpha\beta})^T$, 
$V_{{\bf p},{\bf p}^\prime}^{\alpha\alpha\beta\beta} =  - V_{{\bf p},{\bf p}^\prime}^{\beta\beta\alpha\alpha}$, with ${\bf J}^{\alpha\alpha\beta\beta}$ given by,
\begin{align}
{\bf M}_a^{\alpha\alpha\beta\beta} &=
\frac{1}{2}\left(
\begin{array}{cc}
J_1^a & 0    \\
0 & J_1^a   
\end{array}
\right), \nonumber \\
{\bf M}_b^{\alpha\alpha\beta\beta} &=
\frac{1}{2}\left(
\begin{array}{cccc}
-J_1^b & 0 & 0 & 0   \\
0 & -J_1^{\prime b} & 0 & 0   \\
0 & 0 & 0 & 0   \\
0 & 0 & 0 & 0   \\  
\end{array}
\right), \nonumber \\
{\bf M}_-^{\alpha\alpha\beta\beta} &=
\frac{1}{2}\left(
\begin{array}{cccc}
-J_2^- & 0 & 0 & 0   \\
0 & -J_2^{\prime -} & 0 & 0   \\
0 & 0 & 0 & 0   \\
0 & 0 & 0 & 0   \\  
\end{array}
\right), \nonumber \\
{\bf M}_+^{\alpha\alpha\beta\beta} &=
\frac{1}{2}\left(
\begin{array}{cccc}
-J_2^{\prime +} & 0 & 0 & 0   \\
0 & -J_2^+ & 0 & 0   \\
0 & 0 & 0 & 0   \\
0 & 0 & 0 & 0   \\  
\end{array}
\right),
\end{align}
and $V_{{\bf p},{\bf p}^\prime}^{\alpha\beta\alpha\beta}$, with ${\bf J}^{\alpha\beta\alpha\beta}$ given by,
\begin{align}
{\bf M}_a^{\alpha\beta\alpha\beta} &=
\left(
\begin{array}{cc}
J_1^a & 0    \\
0 & J_1^a   
\end{array}
\right), \nonumber \\
{\bf M}_b^{\alpha\beta\alpha\beta} &=
\frac{1}{2}\left(
\begin{array}{cccc}
J_1^b & 0 & -J_1^b & 0   \\
0 & J_1^{\prime b} & 0 & -J_1^{\prime b}   \\
-J_1^{ b} & 0 & J_1^{ b} & 0   \\
0 & -J_1^{\prime b} & 0 & J_1^{\prime b}   \\  
\end{array}
\right), \nonumber \\
{\bf M}_-^{\alpha\beta\alpha\beta} &=
\frac{1}{2}\left(
\begin{array}{cccc}
J_2^- & 0 & -J_2^- & 0   \\
0 & J_2^{\prime -} & 0 & -J_2^{\prime -}   \\
-J_2^- & 0 & J_2^- & 0   \\
0 & -J_2^{\prime -} & 0 & J_2^{\prime -}   \\  
\end{array}
\right), \nonumber \\
{\bf M}_+^{\alpha\beta\alpha\beta} &=
\frac{1}{2}\left(
\begin{array}{cccc}
J_2^{\prime +} & 0 & -J_2^{\prime +} & 0   \\
0 & J_2^+ & 0 & -J_2^+   \\
-J_2^{\prime +} & 0 & J_2^{\prime +} & 0   \\
0 & -J_2^+ & 0 & J_2^+   \\  
\end{array}
\right),
\end{align}

Expanding the renormalised vertices in the same lattice harmonics gives,
\begin{align}
\Gamma^{\beta\beta\gamma\delta}_{{\bf p},{\bf p}^\prime} = \langle \Gamma^{\beta\beta\gamma\delta}_{{\bf p},{\bf p}^\prime} \rangle +
 \sum_{i=2}^{15}  A^{\gamma\delta}_i  \delta T_i({\bf p}^\prime),
\end{align}
where the $A$ coefficients contain the dependence on ${\bf p}$, ${\bf K}$ and $\Delta$ and,
\begin{align}
\delta T_i({\bf p}) = T_i({\bf p})  - \langle T_i({\bf p})  \rangle.
\end{align}
Substituting the expansions into the set of equations given in Eqs.~\ref{eq:row12}, Eq.~\ref{eq:row3} and Eqs.~\ref{eq:remaining_rows} results in a set of 45 equations.
The first two of these, which descend from Eqs.~\ref{eq:row12} are given by,
\begin{align}
&\langle\Gamma^{\beta\beta\beta\beta}_{{\bf p},{\bf p}^{\prime\prime}} \rangle \int \frac{d^2p^{\prime\prime}}{2\pi^2} 
  g^{\beta\beta}_{{\bf p}^{\prime\prime}}
+\sum_{j=2}^{15}  A_j^{\beta\beta} \int \frac{d^2p^{\prime\prime}}{2\pi^2}  g^{\beta\beta}_{{\bf p}^{\prime\prime}} \delta T_j({\bf p}^{\prime\prime}) \nonumber \\
&+\langle \Gamma^{\beta\beta\alpha\alpha}_{{\bf p},{\bf p}^{\prime\prime}} \rangle \int \frac{d^2p^{\prime\prime}}{2\pi^2}   g^{\alpha\alpha}_{{\bf p}^{\prime\prime}}
+\sum_{j=2}^{15}  A_j^{\alpha\alpha} \int \frac{d^2p^{\prime\prime}}{2\pi^2}  g^{\alpha\alpha}_{{\bf p}^{\prime\prime}} \delta T_j({\bf p}^{\prime\prime})
 =1,
\end{align}
and,
\begin{align}
\langle \Gamma^{\beta\beta\alpha\beta}_{{\bf p},{\bf p}^{\prime\prime}} \rangle \int \frac{d^2p^{\prime\prime}}{2\pi^2} g^{\alpha\beta}_{{\bf p}^{\prime\prime}}
+ \sum_{j=2}^{15}  A_j^{\alpha\beta} \int \frac{d^2p^{\prime\prime}}{2\pi^2} g^{\alpha\beta}_{{\bf p}^{\prime\prime}} \delta T_j({\bf p}^{\prime\prime})
=0.
\end{align}
The third, which comes from Eq.~\ref{eq:row3}, is given by,
\begin{align}
& \langle \Gamma^{\beta\beta\beta\beta}_{{\bf p},{\bf p}^{\prime}} \rangle 
- \langle \Gamma^{\beta\beta\alpha\alpha}_{{\bf p},{\bf p}^{\prime}} \rangle \nonumber \\
&+ \langle \Gamma^{\beta\beta\beta\beta}_{{\bf p},{\bf p}^{\prime\prime}} \rangle \int \frac{d^2p^{\prime\prime}}{2\pi^2}  g^{\beta\beta}_{{\bf p}^{\prime\prime}} (\langle V^{\beta\beta\beta\beta}_{{\bf p}^{\prime\prime},{\bf p}^{\prime}}\rangle - \langle V^{\beta\beta\alpha\alpha}_{{\bf p}^{\prime\prime},{\bf p}^{\prime}}\rangle) \nonumber \\
&+ \langle \Gamma^{\beta\beta\alpha\alpha}_{{\bf p},{\bf p}^{\prime\prime}} \rangle \int \frac{d^2p^{\prime\prime}}{2\pi^2} g^{\alpha\alpha}_{{\bf p}^{\prime\prime}} (\langle V^{\alpha\alpha\beta\beta}_{{\bf p}^{\prime\prime},{\bf p}^{\prime}}\rangle -  \langle V^{\alpha\alpha\alpha\alpha}_{{\bf p}^{\prime\prime},{\bf p}^{\prime}}\rangle) \nonumber \\
&+ \langle \Gamma^{\beta\beta\alpha\beta}_{{\bf p},{\bf p}^{\prime\prime}} \rangle \int \frac{d^2p^{\prime\prime}}{2\pi^2} g^{\alpha\beta}_{{\bf p}^{\prime\prime}} (\langle V^{\alpha\beta\beta\beta}_{{\bf p}^{\prime\prime},{\bf p}^{\prime}}\rangle -  \langle V^{\alpha\beta\alpha\alpha}_{{\bf p}^{\prime\prime},{\bf p}^{\prime}}\rangle) \nonumber \\
&+\sum_{j=2}^{15}  A_j^{\beta\beta}  \int \frac{d^2p^{\prime\prime}}{2\pi^2}  g^{\beta\beta}_{{\bf p}^{\prime\prime}} (\langle V^{\beta\beta\beta\beta}_{{\bf p}^{\prime\prime},{\bf p}^{\prime}}\rangle - \langle V^{\beta\beta\alpha\alpha}_{{\bf p}^{\prime\prime},{\bf p}^{\prime}}\rangle) \delta T_j({\bf p}^{\prime\prime}) \nonumber \\
&+\sum_{j=2}^{15}  A_j^{\alpha\alpha} \int \frac{d^2p^{\prime\prime}}{2\pi^2} g^{\alpha\alpha}_{{\bf p}^{\prime\prime}} (\langle V^{\alpha\alpha\beta\beta}_{{\bf p}^{\prime\prime},{\bf p}^{\prime}}\rangle -  \langle V^{\alpha\alpha\alpha\alpha}_{{\bf p}^{\prime\prime},{\bf p}^{\prime}}\rangle) \delta T_j({\bf p}^{\prime\prime}) \nonumber \\
&+\sum_{j=2}^{15}  A_j^{\alpha\beta}  \int \frac{d^2p^{\prime\prime}}{2\pi^2}  g^{\alpha\beta}_{{\bf p}^{\prime\prime}} (\langle V^{\alpha\beta\beta\beta}_{{\bf p}^{\prime\prime},{\bf p}^{\prime}}\rangle - \langle V^{\alpha\beta\alpha\alpha}_{{\bf p}^{\prime\prime},{\bf p}^{\prime}}\rangle) \delta T_j({\bf p}^{\prime\prime}) \nonumber \\
&= \langle V^{\beta\beta\beta\beta}_{{\bf p},{\bf p}^{\prime}}\rangle - \langle V^{\beta\beta\alpha\alpha}_{{\bf p},{\bf p}^{\prime}}\rangle
\end{align}
The remaining equations are derived from Eqs.~\ref{eq:remaining_rows} and are given by,
\begin{align}
&A_i^{\gamma\delta}
+\langle \Gamma^{\beta\beta\beta\beta}_{{\bf p},{\bf p}^{\prime\prime}} \rangle \int \frac{d^2p^{\prime\prime}}{2\pi^2} g^{\beta\beta}_{{\bf p}^{\prime\prime}}  \sum_{l=1}^{15} T^*_l({\bf p}^{\prime\prime}) J_{li}^{\beta\beta \gamma\delta} \nonumber \\
&+\sum_{j=2}^{15}  A_j^{\beta\beta} \int \frac{d^2p^{\prime\prime}}{2\pi^2} g^{\beta\beta}_{{\bf p}^{\prime\prime}} \delta T_j({\bf p}^{\prime\prime})  \sum_{l=1}^{15} T^*_l({\bf p}^{\prime\prime}) J_{li}^{\beta\beta \gamma\delta}  \nonumber \\
& +\langle \Gamma^{\beta\beta\alpha\alpha}_{{\bf p},{\bf p}^{\prime\prime}} \rangle \int \frac{d^2p^{\prime\prime}}{2\pi^2} g^{\alpha\alpha}_{{\bf p}^{\prime\prime}}  \sum_{l=1}^{15} T^*_l({\bf p}^{\prime\prime}) J_{li}^{\alpha\alpha \gamma\delta}  \nonumber \\
&+\sum_{j=2}^{15}  A_j^{\alpha\alpha} \int \frac{d^2p^{\prime\prime}}{2\pi^2} g^{\alpha\alpha}_{{\bf p}^{\prime\prime}} \delta T_j({\bf p}^{\prime\prime}) \sum_{l=1}^{15} T^*_l({\bf p}^{\prime\prime}) J_{li}^{\alpha\alpha \gamma\delta} 
\nonumber \\
& +\langle \Gamma^{\beta\beta\alpha\beta}_{{\bf p},{\bf p}^{\prime\prime}} \rangle \int \frac{d^2p^{\prime\prime}}{2\pi^2} g^{\alpha\beta}_{{\bf p}^{\prime\prime}}  \sum_{l=1}^{15} T^*_l({\bf p}^{\prime\prime}) J_{li}^{\alpha\beta \gamma\delta}  \nonumber \\
&+\sum_{j=2}^{15}  A_j^{\alpha\beta} \int \frac{d^2p^{\prime\prime}}{2\pi^2} g^{\alpha\beta}_{{\bf p}^{\prime\prime}} \delta T_j({\bf p}^{\prime\prime}) \sum_{l=1}^{15} T^*_l({\bf p}^{\prime\prime}) J_{li}^{\alpha\beta \gamma\delta} 
\nonumber \\
&= \sum_{l=1}^{15} T^*_l({\bf p}) J_{li}^{\beta\beta \gamma\delta},
\end{align}
which encodes 14 equations for each $\gamma\delta \in \{ \beta\beta,\alpha\alpha, \alpha\beta \}$.

These 45 equations can be used to generate a matrix equation in which a matrix multiplies the vector,
\begin{align}
 \left( 
\langle \Gamma^{\beta\beta\beta\beta}_{{\bf p},{\bf p}^{\prime\prime}} \rangle, {\bf A}^{\beta\beta},
\langle \Gamma^{\beta\beta\alpha\alpha}_{{\bf p},{\bf p}^{\prime\prime}} \rangle, {\bf A}^{\alpha\alpha},
\langle \Gamma^{\beta\beta\alpha\beta}_{{\bf p},{\bf p}^{\prime\prime}} \rangle, {\bf A}^{\beta\alpha}
 \right),
\end{align}
where the ${\bf A}^{\gamma\delta}$ are 14-component vectors made up of the $A_i^{\gamma\delta}$ terms.
When the matrix that multiplies this vector has a zero determinant, it implies that at least one of the components of the vector diverges, and therefore that one of the $\Gamma$ vertices diverges, implying the existence of a bound state.
Since the $\beta$ bosons have a gapless $\epsilon_{\bf q}^\beta$, it must be $\Gamma^{\beta\beta\beta\beta}_{{\bf p},{\bf p}^{\prime\prime}}$ that diverges.

\bibliographystyle{apsrev4-1}
\bibliography{bibfile}


\end{document}